\documentclass[a4paper, 11pt]{article}
\usepackage{latexsym,graphics,subfigure}  
\usepackage[dvips]{graphicx}
\begin{document}

\title{\bf{Three-dimensional lattice-Boltzmann simulations of critical
spinodal decomposition in binary immiscible fluids}}

\author{ N\'elido Gonz\'alez-Segredo
\footnote{\small  Also at Grup de F\'\i sica Estad\'\i stica,
Universitat Aut\`onoma de Barcelona, 08193 Bellaterra, Barcelona,
Spain.  \normalsize }\\ 
\small{\em Centre for Computational Science}			\\
\small{\em Department of Chemistry, University College London}	\\
\small{\em 20 Gordon Street, London WC1H 0AJ, UK} 		\\
\small{\texttt{n.gonzalez-segredo@ucl.ac.uk}}			\\
\vspace{0.1cm}							\\ 
Maziar Nekovee							\\ 
\small{\em Complexity Research Group, BT Laboratories}		\\ 
\small{\em Martlesham Heath, Ipswich, Suffolk IP5 3RE, UK}	\\
\small{\texttt{maziar.nekovee@bt.com}}				\\ 
\normalsize
\vspace{0.1cm}							\\ 
and 								\\
\vspace{0.1cm}							\\ 
Peter V. Coveney						\\
\small{\em Centre for Computational Science}			\\
\small{\em Department of Chemistry, University College London}  \\
\small{\em 20 Gordon Street, London WC1H 0AJ, UK}		\\
\small{\texttt{p.v.coveney@ucl.ac.uk}}				\\ 
\normalsize
}
%\address{} \date{\today}
\maketitle
%\tableofcontents
\newpage

\begin{abstract}
\noindent We use a modified Shan-Chen, noiseless lattice-BGK model for
binary immiscible, incompressible, athermal fluids in three
dimensions to simulate the coarsening of domains following a deep
quench below the spinodal point from a symmetric and homogeneous
mixture into a two-phase configuration. The model is derivable from a
continuous-time Boltzmann-BGK equation in the presence of an
intercomponent body force. We find the average domain size growing
with time as $t^\gamma$, where $\gamma$ increases in the
range $0.545\pm 0.014<\gamma<0.717\pm 0.002$, consistent with a
crossover between diffusive $t^{1/3}$ and hydrodynamic viscous,
$t^{1.0}$, behaviour. We find good collapse onto a single scaling 
function, yet the domain growth exponents differ from others' works'
for similar values of the unique characteristic length $L_0$ and time
$T_0$ that can be constructed out of the fluid's parameters. This
rebuts claims of universality for the dynamical scaling
hypothesis. For $\mathrm{Re}=2.7$ and small wavenumbers, $q$, we also
find a $q^2\leftrightarrow q^4$ crossover in the scaled structure
function, which disappears when the dynamical scaling reasonably
improves at later stages ($\mathrm{Re}=37$). This excludes noise as
the cause for a $q^2$ behaviour, as analytically derived from Yeung
and proposed by Appert {\it et al.} and Love {\it et al.} on the basis
of their lattice-gas simulations. We also observe exponential temporal
growth of the structure function during the initial stages of the
dynamics and for wavenumbers less than a threshold value, in
accordance with the diffusive Cahn-Hilliard Model B. However, this
exponential growth is also present in regimes proscribed by that
model. There is no evidence that regions of parameter space for which
the scheme is numerically stable become unstable as the simulations
proceed, in agreement with finite-difference relaxational models and
in contradistinction with an unconditionally unstable lattice-BGK
free-energy model previously reported. Those numerical instabilities
that do arise in this model are the result of large intercomponent
forces which turn the equilibrium distribution negative. 
\end{abstract}

\newpage

\section{Introduction}

Homogeneous binary fluid mixtures phase segregate into two phases with
different compositions when quenched into thermodynamically unstable
regions of their phase diagram, a process also called spinodal
decomposition. This is achieved by lowering the temperature well below
the so called spinodal temperature. For incompressible, 50:50
mixtures, also called critical or symmetric mixtures, these phases
form interconnected domains which at late times produce a bicontinuous
structure with sharp, well developed interfaces. For asymmetric
mixtures (phases with different densities) there is a phase transition
at early times from an interpenetrating structure termed as
``bicontinuous'' to the so-called ``droplet phase'', which in turn
undergoes subsequent coarsening via coalescence \cite{PERROT94}. The
composition of a binary immiscible fluid is one of the variables
affecting its dynamics. Fields where spinodal decomposition is of
industrial relevance comprise the metallurgical, oil, food, paints and
coatings industries. Polymer blends and gels immersed in a solvent are
also potentially important applications where phase separation occurs
and needs to be controlled \cite{GUNTON,BALAZS}.

Spinodal decomposition has been extensively studied by experimental
\cite{PERROT99}, analytical \cite{GRANT85,SOLIS00}, and numerical
\cite{CHAKRA89,VALLS,LOVE,EMERTON97,KENDON99,WAGNER98,JURY99,MD-SD,
PAGONA01,PAGONABARRAGA-NJP} approaches. The fact that it entails a
variety of mechanisms that can act concurrently and at different
length and time scales has made it a testbed for complex fluid
simulation methods. Among the latter are hydrodynamic lattice gases
\cite{LGA}, the lattice Boltzmann equation \cite{LBE-REV}, and
dissipative particle dynamics \cite{DPD}.

Despite all the interest attracted by the subject, how the mechanisms
responsible for domain separation act remains on unsettled grounds. In
particular, the dynamics of the late time, true asymptotic growth is
unclear. Also, the dynamical scale invariance hypothesis (to be
explained later on in this paper), which is treated almost as
canonical by analytical and numerical approaches to solving the
continuum, local-thermodynamic Cahn-Hilliard equations, has
experimentally been proven to fail at least under certain conditions
\cite{TANAKA}. 

Numerical studies on spinodal decomposition include methods at the
{\em macroscopic scale}, based on the numerical solution of either the
Navier-Stokes \cite{NAVSTOK,TRYGGVASON} or the Cahn-Hilliard equations
\cite{CHC,CHAKRA89,FARRELL89}, the {\em mesoscopic scale}, where
lattice-Boltzmann (LB) methods \cite{SHAN-CHEN,OXFORD-MODEL}, lattice
gases \cite{EMERTON97,LOVE}, dissipative particle dynamics
\cite{DPD-SD}, and Ising \cite{FRATZL} approaches are examples, and
the {\em microscopic scale}, with classical molecular dynamics
\cite{MD-SD}.

Fluid dynamical methods in the mesoscopic scale came to light as a way
to grasp  the relevant thermohydrodynamical behaviour with as little
computational effort as possible. This is achieved by evolving a
microworld in which the usual vast number of molecular degrees of
freedom and characterisation have been drastically reduced, based on
the fact that, away enough from critical points, a fluid's macrostate
is pretty much insensitive to many of its microscopic properties.
Some regard the Cahn-Hilliard equations to be within the mesoscopic
scale. They derive from the van der Waals' formulation of quasilocal
thermodynamics \cite{ROWLINSON}, extended by Cahn and Hilliard
\cite{CHC}, and aim at solving a Langevin-like diffusion equation for
the conserved order parameter. This equation involves a chemical
potential derived from a phenomenological, Ginzburg-Landau expansion
for the free energy, and leads to phase segregation if the temperature
is below a critical value. The scheme commonly used for the study of 
phase segregation in immiscible fluids is termed Cahn-Hilliard {\it
Model H} \cite{HOHENBERG}; hydrodynamics is included by introducing
mass currents, which couple the diffusion equation with the
Navier-Stokes equation. Thermal effects are sometimes included in the
dynamics by the addition of a noise term satisfying a
fluctuation-dissipation theorem. 

Cahn-Hilliard equations have been applied to model the segregation
dynamics of deep and sudden thermal quenches of fluid mixtures. Such
quenches are usually chosen to be sudden to avoid thermal noise
effects and set up an initial condition that quickly leads to a state
of steep domain walls and where diffusion is negligible compared with
hydrodynamic effects, thus leaving the conditions that the dynamical 
scaling hypothesis requires. However, local-equilibrium cannot be
guaranteed for a mixture undergoing a sudden quench, which puts the
existence of a free energy and the equilibrium states modelled by it
on rather shaky grounds.

The lattice-Boltzmann method we use in this work is the Shan-Chen
lattice-BGK scheme for binary immiscible and incompressible fluid flow
\cite{SHAN-CHEN}.  The equilibrium state for each pure fluid is chosen
to be a local isothermal Maxwellian, and Shan and Chen's
contribution to the lattice-BGK scheme comes through the phase
separation prescription. This is incorporated via intercomponent
repulsive mean-field forces between elements of fluids (meant to be at
a mesoscopic scale) which alter the local equilibrium, and not through
a local equilibrium reproducing a chemical potential derived from a 
free-energy functional. The Shan-Chen method has been used by Martys
and Douglas \cite{MARTYS-DOUGLAS} to qualitatively simulate  spinodal 
decomposition for critical and off-critical quenches in 3D. There have
been recent quantitative studies in 2D using this method for critical
spinodal decomposition \cite{MAZIAR-ET-AL,CHIN01}. An early study on
critical 2D and 3D spinodal decomposition was put forward by Alexander
{\em et al.} \cite{ALEXANDER93} using the lattice-Boltzmann method
proposed by Gunstensen {\em et al} \cite{GUNSTENSEN91}.

Lattice-BGK methods based on a Ginzburg-Landau free-energy functional
\cite{OXFORD-MODEL} achieve multiphase behaviour by using two separate
distribution functions: one for the mass density and one for the order
parameter, this being defined as the difference between the phases'
densities. Higher-order velocity moments of these distributions
are imposed to coincide with thermomechanical quantities obtained from 
the free energy. The term ``top-down'' is used in the literature to
address this type of approach, whereas we shall use ``bottom-up'' in
the remainder to signify fully mesoscopic methods. Some criticisms of
top-down approaches \cite{LUO} include their frequent lack of Galilean  
invariance (although Inamuro \cite{INAMURO} presented a model that
does exhibit this property), and their phenomenological
character. Studies of spinodal decomposition using these methods are
described in the works of Wagner {\em et al.}
\cite{WAGNER98,WAGNER-YEOMANS}, Kendon {\em et al.} \cite{KENDON99},
and Cates {\em et al.} \cite{CATES99}. 

Numerical instabilities are a great cause for concern in
lattice-Boltzmann methods, a study of which will be addressed for the 
lattice-BGK method we employ in this work. Their sources are two-fold:
(a) the finite-difference, discrete-velocity scheme used to solve the 
BGK-Boltzmann equation prevents the existence of an H-theorem, and (b)
the  approximations used for the equilibrium distribution do not
guarantee its positivity, and hence that of the nonequilibrium 
distribution. Linear stability analyses have been applied to the
lattice-BGK model by Sterling \cite{STERLING}, and in more detail by
Lallemand and Luo \cite{LALLEMAND-LUO} comparing a lattice-BGK model
to a generalised LB model with a different relaxation time for each
physical flux. Qian {\em et al.} \cite{QIAN} gave conditions for the
Mach number and the shear viscosity such that the lattice-BGK scheme
produces positive mass densities. New approaches to unconditionally
stable lattice-Boltzmann models have recently appeared too
\cite{KARLIN,CHEN-TEIXEIRA,BOGHOSIAN-LBE,ENTROPIC-LBE-PVC}. They prove
the existence of functionals satisfying an H-theorem.

Our objective in this work is to present a bottom-up lattice-BGK
method for the study of scaling laws in the spinodal decomposition of
critical fluid mixtures in three dimensions. This method has certain
advantages over lattice-BGK methods based on a free-energy functional,
namely, a smaller number of free parameters to tune, Galilean
invariance guaranteed, and a simpler equilibrium
distribution. Moreover, it refuses to inject macroscopic information
into the mesoscopic dynamics as the top-down methods do, on the
grounds that for lattice-BGK methods there is no H-theorem available
that guarantees an unconditional approach to a given equilibrium.
Indeed, in the context of general complex fluid applications, an
expression for the free energy itself may be unknown, and/or its
validity be questioned for regimes far enough from local equilibrium,
making a top down approach not even viable.

The remainder of the paper is structured as follows. In Section
\ref{SPIN-DECOMP} we discuss the dynamical scaling hypothesis, which
asserts that after the quench all length scales in the mixture share
the same growth law with time. The modified Shan-Chen model we use is
explained  in Section \ref{SCLB}; Section \ref{SURFTENS} introduces
the method we use to measure surface tension, while in Section
\ref{SIMUL} we describe the simulations performed and the growth laws
and scaling functions drawn from them which allow to test the
validity of the dynamical scaling hypothesis. Finally, we present our
Conclusions in Section \ref{CONCLU}.

\section{Spinodal decomposition}  
\label{SPIN-DECOMP}

After domain walls have achieved their thinnest configuration via
diffusion, the time evolution of the bicontinuous structure that is
produced in the phase segregation process that symmetric mixtures
undergo presents geometrical self-similarity to the initial stages of
such a process when the structure is zoomed in at increasing
magnification. This leads us to the {\em dynamical scaling
hypothesis}, which states that at late times, when diffusive effects
have died out, there is a unique characteristic length scale $L$ which
grows with time such that the geometrical structure of domains is (in
a statistical sense) independent  of time when lengths are scaled by $L$
\cite{BRAY}. This amounts to saying that all length scales have the
same time evolution. Such a characteristic length scale must be
universal for all fluids with the same shear viscosity, $\eta$,
density, $\rho$, and surface tension $\sigma$, provided that no
mechanisms are involved in their late stage growth other than viscous
dissipation, fluid inertia and capillary forces, respectively. This is
so because, as we shall see later on, only one length scale can be
constructed out of the fluid's parameters $\eta$, $\rho$, and
$\sigma$, these being the only ones present in a hydrodynamic
description of the mixture via the Navier-Stokes equations.

The characteristic length scale is usually measured by looking at the
first zero crossing of the equal-time pair-correlation function of the
order parameter fluctuations \cite{GUNTON},

\begin{equation}
	C({\mathbf r},T)
	\equiv
	\langle
	\phi^\prime(\mathbf{x+r},T)\phi^\prime({\mathbf x},T)
	\rangle\,,  \label{PAIR_CORR_FUNCT}
\end{equation}

\noindent where, on the lattice, $\langle\rangle\equiv\sum_{\mathbf
x}\varsigma/V$, $V$ is the spatial volume, $\varsigma$ is the volume
of the lattice's unit cell (hence $V/\varsigma$ is the number of nodes
in the lattice), $T$ is the time parameter in time steps, ${\mathbf
r}$ and ${\mathbf x}$ are spatial vectors, and
$\phi^\prime\equiv\phi-\langle\phi\rangle$ are the order parameter
fluctuations, where $\phi({\mathbf x}) \equiv \rho^{\mathrm
R}({\mathbf x})-\rho^{\mathrm B}({\mathbf x})$ is the order parameter
for our binary fluid (say, a mixture of red (R) and blue (B) 
phases). The units of $C({\mathbf r},T)$ are squared mass density. In
the remainder, ``lattice units'' will mean unity for the mass, length,
and time units, respectively, in an arbitrary unit system. The Fourier
transform of $C({\mathbf r},T)$, called the structure function, is  

\begin{equation}
	S({\mathbf k},T)
	=
	\frac{\varsigma}{V}\Big|\phi_{\mathbf
	k}^\prime(T)\Big|^2		\,. \label{STRUCT_FUNCT}
\end{equation}

\noindent The units for the structure function are the same as those
for the correlation function, and $\phi_{\mathbf k}^\prime$ is the
Fourier transform of the fluctuations. Function (\ref{STRUCT_FUNCT})
is volume-normalised, and gives no power spectrum for infinite
lengths, i.e. 

$$
	\frac{\varsigma '}{V'} \sum_{\mathbf k}S({\mathbf k},T)=1	
	\,,\qquad
	S({\mathbf k=0},T)=0.
$$

\noindent where $\varsigma '$ is the unit cell volume in reciprocal
space, and $V'=(2\pi/L)^3 V/\varsigma=(2\pi)^3/\varsigma$ is the
reciprocal space volume; in fact,
$\varsigma'/V'=\varsigma/V$. Although Eqs. (\ref{PAIR_CORR_FUNCT}) and 
(\ref{STRUCT_FUNCT}) are numerically equivalent, the intensity of
X-ray and neutron scattering is directly proportional to the structure 
function, which is hence easily measurable; it is thus this quantity
that we prefer to use to measure the system's characteristic length
scales.

We define the (time-dependent) characteristic size $L$ of the domains
as    

\begin{equation}
	L(T) \equiv \frac{2\pi}{k_1(T)}\,, \label{DOMAIN-SIZE}
\end{equation}

\noindent in lattice units, where $k_1(T)$ is the first moment (mean),

\begin{equation}
	k_1(T) \equiv \frac{\sum_k kS(k,T)}{\sum_kS(k,T)}	\,,
	\label{FIRST_MOMENT}
\end{equation}

\noindent of the spherically averaged structure function $S(k,T)$,
defined by

\begin{equation}
	S(k,T) \equiv \frac{\sum_{\mathbf {\hat k}}S({\mathbf k},T)}
	{\sum_{{\mathbf{\hat k}}} 1}				\,,
	\label{SASF}
\end{equation}

\noindent where ${\mathbf {\hat k}}$ indicates the set of wave vectors
contained in a spherical shell of thickness one (in reciprocal-space
lattice units) centered around ${\mathbf k}$, i.e.  such that
$n-\frac{1}{2}\le \frac{V^{1/3}}{2\pi}k\le n+\frac{1}{2}$,
$n$  being an integer. $k$ is the modulus of ${\mathbf  k}$ which is
smaller than the Nyquist critical frequency $k_c=\pi$ to prevent
aliasing. In the limit of short distances and large momenta, scaling
arguments lead \cite{BRAY} to the relation 

\begin{equation}
	S({\mathbf k},T)\sim\frac{1}{Lk^{D+1}} \label{POROD}
\end{equation}

\noindent valid for $kL\gg 1$, also known as Porod's law, where $D$ is
the spatial dimension. Short distances here means $\xi\ll r \ll L$,
where $\xi$ is the interface thickness. 

Other measures have also been used for the system's characteristic
length scale, namely, the position of the structure function's
maximum, and the structure function's second moment, $k_2$
\cite{GUNTON}. We chose to use the first moment $k_1$ as it is the
simplest quantity among the aforementioned. Appert {\it et al.}
\cite{APPERT-LGASD} found that the structure function's maximum's
wavenumber provided a length evolving similarly, although in a noisier 
fashion, to that derived from the first moment. 

Mathematically, the dynamical scaling hypothesis can be written as

\begin{equation}
	C({\mathbf r},T)=f(r/L)				\,, 
	\label{SCALING-CORR}
\end{equation}

\noindent or

\begin{equation}
	S({\mathbf k},T) = L^Dg(kL)			\,, 
	\label{COLLAPSED-SF}
\end{equation}

\noindent where $L=L(T)$ is a function of time, and $g$ is the Fourier
transform of $f$, both of which are the same for any late time slice.
  
Using methods introduced by Kendon {\em et al.} \cite{KENDON99}, there
are unique length and time units that can be defined from the fluid's
density, shear viscosity and surface tension, $\rho$, $\eta$, and
$\sigma$, respectively, as follows: 

\begin{eqnarray}
	L_0 \equiv \frac{\eta ^2}{\rho \sigma}		\,&,&\qquad
	T_0 \equiv \frac{\eta ^3}{\rho \sigma ^2} 	\,.
	\label{L0T0}
\end{eqnarray}

\noindent We can think of these as a wavelength and a period
associated with the system's fluctuations, respectively, although they
do not necessarily have to refer to actual fluctuation averages. We
can define the dimensionless variables: 

\begin{equation}
	l  \equiv L/L_0					\,,\qquad 
	t  \equiv (T-T_{\mathrm{int}})/T_0		\,,
\end{equation}

\noindent which serve to express the universal character
of the dynamical scaling hypothesis. Parameter $T_{\mathrm{int}}$ is
an offset that allows one to account for early time diffusional
transients and lattice effects. Due to the finite resolution of the
lattice the initial condition is not an infinitely fine-grained
thorough mixture ($\phi=0$) but there is a non-negligible domain size
measured at time $T=0$. We have then to specify a time origin prior to
$T=0$, corresponding to a fictitious zero domain size. 

For a critical binary immiscible mixture in three dimensions, scaling
arguments applied to the terms of the Cahn-Hilliard Model-H equations
show that Eq. (\ref{SCALING-CORR}) holds in the asymptotic limit
\cite{BRAY}, or, equivalently, that

\begin{equation}
	l \propto t^\gamma				\,,
\label{EXPONENTS}
\end{equation}

\noindent where $\gamma=1$ and $\gamma=2/3$ for the cases when
hydrodynamic viscosity and inertia dominate the dynamics,
respectively. From the Cahn-Hilliard Model B, which is a 
Langevin diffusion equation without noise conserving the order
parameter \cite{HOHENBERG}, an exponent of $\gamma=1/3$ is derived,
identical to that obtained from the Lifshitz-Slyozov theory for the
growth of a minority phase whose volume fraction is
negligible, and is expected to appear at diffusive stages, before
hydrodynamics kicks in. Scaling theories do not give any prediction
for the crossovers' positions in time other than that they are ``of
order unity'' \cite{KENDON01}.   

Using a free-energy based, lattice-BGK method, Cates {\em et al.}
\cite{CATES99} reached the viscous regime ($l \propto t$) for $L_0
\approx 5.9$ and $\textrm{Re} < 0.1$, and the inertial regime ($l
\propto t^{2/3}$) for $L_0 \approx 0.0003$ and $\textrm{Re} <
350$. The Reynolds number is defined in this domain-coarsening context
as  

\begin{equation}
	\textrm{Re}\equiv
	\frac{L}{\nu}\frac{\mathrm{d}L}{\mathrm{d{T}}}=
	l\dot{l}					\,.
\end{equation}

\noindent where $\nu$ is the kinematic viscosity of the fluid mixture,
as defined in the next section, and $\dot{l}$ is the time derivative 
$\mathrm{d}l/\mathrm{d}t$. 

There is also experimental \cite{TANAKA} and 2D simulation
\cite{WAGNER98} evidence of breakdown of scale invariance in symmetric
binary immiscible quenches. In those experiments, the breakdown of
scale invariance occurs \cite{TANAKA} for  symmetric binary mixtures
in confined geometries under the influence of wetting, and a
universality has been reported to hold. The process consists of a
hydrodynamic coarsening occurring faster than mass diffusion, leaving
the  system with macroscopic domains whose concentrations are near to
but not at the coexisting equilibrium ones. Metastability or
instability of the domains then causes a secondary phase separation 
to kick in via diffusion. Scale invariance and self-similarity have
also been recently found to break down for viscoelastic binary fluid 
mixtures \cite{ARAKI}. Finally, there is simulation evidence of
breakdown of scale invariance coming from free-energy based,
lattice-BGK simulations in 2D. The rationale for this is the
coexistence of competing mechanisms at all times in the
mixture: diffusion, hydrodynamic modes, and surface tension, giving
rise to length scales with different growth exponents
\cite{WAGNER98}.

\section{Our lattice-Boltzmann model}
\label{SCLB}

Initially introduced as a coarse grained version of the lattice-gas
automaton method for fluid flow simulation, the lattice-Boltzmann
model can also be interpreted as a finite difference solver for
the Bhatnagar-Gross-Krook (BGK) approximation to the Boltzmann
transport equation \cite{LBE-REV}. From lattice gases it inherits a
particulate, mesoscopic character, as their particles can be
assimilated to any physical size which is negligible at a hydrodynamic
scale; moreover, unlike lattice-gas automata, no fluctuations are
present within the scheme \cite{LBE-FLUCT}. From the simplicity of the
Boltzmann-BGK collision term the LB method gains algorithmic
efficiency in simulating fluid flow over solving the incompressible
Navier-Stokes equations. When extended to multiphase flows, these
features are  especially valuable in looking at the complicated domain
interfaces that arise in the coarsening of binary mixtures.

The method we use is a modification of the multicomponent, immiscible
fluid LB scheme of Shan and Chen \cite{SHAN-CHEN}, which will be
explained in detail in Subsection
\ref{MODIFICATION-EXPLAINED-HERE}. The Shan-Chen LB model employs an
expansion in Mach number of a  Maxwellian equilibrium
distribution. Phase-segregating interactions are introduced by means
of a self-consistently generated mean-field force between
particles. The inclusion of this force gives rise to a non-ideal gas
equation of state through the Navier-Stokes equation, which is
reproduced via the usual multiscale Chapman-Enskog \cite{FERZIGER} or
moment (Grad)  \cite{GRAD} expansion of the distribution function. No
thermohydrodynamic behaviour is imposed on the equilibrium
distribution, as aforementioned free-energy based, lattice-BGK methods
do \cite{OXFORD-MODEL}, partly because none of the lattice-Boltzmann
implementations reported in the literature so far exhibit an H-theorem
ensuring the existence of an asymptote towards a prescribed
equilibrium, and partly because a purely mesoscopic, mean-field
approach is preferred here. The coefficients of the equilibrium
distribution expansion are determined by the conservation of mass and
momentum, the property that Galilean invariance holds, and an
isotropic pressure tensor.

In this work we employ a pseudo four-dimensional lattice, which is the
projection onto 3D of the D4Q24 face-centered hypercubic (FCHC),
single-speed lattice, where the notation implies the spatial dimension
(4) and the number of vectors linking a site to its nearest neighbours
(24). The FCHC lattice guarantees isotropic behaviour for the
macrosopic momentum balance equation \cite{FRISCH-ET-AL}.

In the following subsections we introduce our modified Shan-Chen
model, first by looking at a non-interacting mixture of gases, and
second including the mean-field force term which gives rise to a
non-ideal gas equation of state. Then, we modify the collision term
such that the Shan-Chen scheme is consistent with that derived from a
Boltzmann-BGK equation in the presence of a force.

\subsection{Mixture of ideal gases}
\label{IDEAL_GASES}

The finite-difference, finite-velocity fully-Lagrangian
\cite{STERLING} scheme for the numerical solution of the
multicomponent Boltzmann equation,

\begin{equation}
	n_k^\alpha(\mathbf{x + c}_k,t+1) - n_k^\alpha({\mathbf x},t)
	= \Omega_k^\alpha \,,
\label{BGK-OP-NOFORCE}
\end{equation}

\noindent governs the time evolution of the $k$th velocity's particle
number density $n_k^\alpha$ for the fluid species $\alpha$ in a
non-interacting mixture of gases. The lattice-BGK collision term is

\begin{equation}
	\Omega_k^\alpha({\mathbf x},t) 
	\equiv 
	-\frac{n_k^\alpha({\mathbf x},t) 
	- n_k^{\alpha(\mathrm{eq})}({\mathbf x},t)}
	{\tau^\alpha}\,,
\end{equation}

\noindent where the time increment and lattice spacing are both unity,
${\mathbf c}_k$ is one of the 24 discrete velocity vectors 
plus one  null velocity, ${\mathbf x}$ is a point of the underlying
Bravais lattice, and $\alpha=\mathrm{R},\mathrm{B}$ (e.g. oil (R) or
water (B)). The parameter $\tau^\alpha$ defines a single relaxation
rate towards equilibrium for component $\alpha$. The function
$n_k^{\alpha(\mathrm{eq})}({\mathbf x},t)$ is the discretisation of a
third-order expansion in Mach number of a local Maxwellian
\cite{HE-LUO},

\begin{equation}
	n_k^{\alpha(\mathrm{eq})}({\mathbf x},t)
	= 
	\omega_k \, n^\alpha({\mathbf x},t)
	\Big[ 1 +
	\frac{1}{c_s^2}{\mathbf c}_k\cdot{{\mathbf u}} +
	\frac{1}{2c_s^4}({\mathbf c}_k\cdot{{\mathbf u}})^2 -
	\frac{1}{2c_s^2}u^2 + \frac{1}{6c_s^6}({\mathbf
	c}_k\cdot{{\mathbf u}})^3 - \frac{1}{2c_s^4}u^2({\mathbf
	c}_k\cdot{{\mathbf u}}) \Big]\,,
\label{EQUIL}
\end{equation}

\noindent where $\omega_k$ are the coefficients resulting from the
velocity space discretisation, and  $c_s$ is the
speed of sound, both of which are determined by the choice of the
lattice. For the projected-D4Q24 lattice we use, the speed of sound is
$c_s=1/\sqrt{3}$, and $\omega_k=1/3$ for the speed $c_k=0$ and $1/36$
for speeds $c_k=1,\sqrt{2}$ \cite{QIAN}. (The projection from 4D to 3D
puts an additional speed into play, $\sqrt{2}$.) In
Eq. (\ref{EQUIL}), ${\mathbf u}$ is the macroscopic velocity of the
mixture, through which the collision term couples the different
velocities ${\mathbf c}_k$, and is a function of ${\mathbf x}$ and
$t$. Also, $n^\alpha({\mathbf x},t)$ is the local particle density for
the $\alpha$-th component, defined as $\sum_k n_k^\alpha({\mathbf
x},t)$.

The judicious choice of the coefficients in the expansion of the
equilibrium distribution (\ref{EQUIL}) allows for mass and momentum to 
be conserved,

\begin{equation}
	\sum_k\Omega_k^\alpha=0		\,,\qquad		
	\sum_\alpha m_\alpha \sum_k{\mathbf
	c}_k\Omega_k^\alpha=0		\,. 	 
	\label{BGK-CONSERVATIONS}
\end{equation}

\noindent Momentum conservation requires the fluid's  macroscopic
velocity ${\mathbf u}$ to be defined in terms of the macroscopic
velocity ${\mathbf u}^\alpha$ for component $\alpha$, 

\begin{equation}
	n^\alpha({\mathbf x},t)\,{\mathbf u}^\alpha 
	\equiv 
	\sum_k n_k^\alpha({\mathbf x},t)\,{\mathbf c}_k 	\,,
	\label{SPECIES_MACR_VEL}
\end{equation}

\noindent as the solution of the three equations:

\begin{equation}
	{\mathbf \Xi}({\mathbf u})  
	= 
	{\mathbf v}			\,,
	\label{FLUID_VELOCITY} 
\end{equation}

\noindent where

\begin{equation}
	\Xi_i({\mathbf u}) \equiv (2-3u^2)u_i+3u_i^3 + 3u_i u_{i+1}^2
	+ 3u_i u_{i+2}^2		\,,
\end{equation}

\noindent with the cartesian index $i$ ranging in the $i\mathrm{mod}3$
set, and ${\mathbf v}$ being defined as the special average:

\begin{equation}
	{\mathbf v} 
	\equiv
	\sum_\alpha\frac{\rho^\alpha{\mathbf u}^\alpha} {\tau^\alpha}
	/ \sum_\alpha\frac{\rho^\alpha}
	{\tau^\alpha}			\,.
\end{equation}

Based on previous experience with lower orders, our choice of a
third-order Taylor expansion in Mach number for the Maxwellian
equilibrium distribution is an attempt to improve the approximation
for velocities which, within the incompressibility limit, are large
enough to make either the distribution funtion become negative or the
error in the expansion too large.

\subsection{Mixture of interacting, non-ideal gases}
\label{MODIFICATION-EXPLAINED-HERE}

In order to deal with non-ideal gases, in particular, fluid mixtures
whose volume elements interact among themselves, each fluid is forced
to relax to a local equilibrium which is modified by the presence of
its surrounding volume elements. The mean-field force density felt by
phase $\alpha$ at site ${\mathbf x}$ and time $t$ from its
surroundings is defined as:

\begin{equation}
	{\mathbf F}^{\alpha}({\mathbf x},t)
	\equiv
	-\psi^\alpha({\mathbf
	x},t)\sum_{\bar\alpha} g_{\alpha\bar\alpha} \sum_{{\mathbf
	x}'}  \psi^{\bar\alpha}({\mathbf x}',t) ({\mathbf x}'-{\mathbf
	x}) \label{LBE-FORCE}
\end{equation}

\noindent where $g_{\alpha\bar\alpha}$ ($>0$ for immiscible fluids) is
a coupling matrix whose non-diagonal elements control interfacial
tension, and $\psi^\alpha$ is the so-called {\em effective mass},
which serves as a functional parameter and can have a general form for
modelling various types of fluids. For simplicity in our
implementation, we have chosen $\psi^\alpha({\mathbf x},t)\equiv
n^\alpha({\mathbf x},t)$   \cite{MAZIAR-ET-AL} and only allowed
nearest-neighbour interactions, ${\mathbf x}'\equiv {\mathbf
x}+{\mathbf c}_k$. Other choices for $\psi$ have also been made
\cite{SHAN-CHEN}.

Shan and Chen \cite{SHAN-CHEN} incorporated the above force term in
the collision substep of the LB dynamics by adding the increment

\begin{equation}
	\Delta{\mathbf u}^\alpha  
	\equiv 
	\frac{{\mathbf
	F}^\alpha}{\rho^\alpha}\tau^\alpha
\end{equation}

\noindent to the velocity ${\mathbf u}$ which enters the second-order
expansion of the equilibrium distribution function. We perform the
same procedure for our third-order expansion (\ref{EQUIL}), obtaining
additional terms:

\begin{eqnarray}
	n_k^{\alpha\mathrm{(eq)}}
	({\mathbf u}+\Delta{\mathbf u}^\alpha) 
	& = &
	n_k^{\alpha\mathrm{(eq)}}({\mathbf u})		\nonumber\\ 
	& + & 
	\omega_k 
	n^\alpha\bigg[  
	\frac{{\mathbf c}_k-{\mathbf u}}{c_s^2} +
	\frac{(2{\mathbf c}_k\cdot{\mathbf u}-u^2)}{2c_s^4}{\mathbf
	c}_k \bigg]\cdot{\mathbf a}^\alpha\tau^\alpha	\nonumber\\ 
	& + & 
	\frac{1}{2}\omega_k n^\alpha\bigg[
	\frac{{\mathbf a}^\alpha\cdot{\mathbf a}^\alpha}{c_s^2} -
	\frac{({\mathbf c}_k\cdot{\mathbf a}^\alpha)^2}{c_s^4} +
	\frac{({\mathbf c}_k\cdot{\mathbf u}) ({\mathbf
	c}_k\cdot{\mathbf a}^\alpha)^2}{c_s^6}
	\bigg](\tau^\alpha)^2				\nonumber\\ 
	& + &
	\frac{1}{6c_s^6}\omega_k n^\alpha 
	({\mathbf c}_k\cdot{\mathbf a}^\alpha\tau^\alpha)^3	\,.  
	\label{SHAN-CHEN-EQDIST}
\end{eqnarray}

\noindent where ${\mathbf a}^\alpha \equiv {\mathbf
F}^\alpha/\rho^\alpha$. 

Luo \cite{LUO} and Martys {\em et al.} \cite{MARTYS} expanded both the
velocity space gradient in the BGK-Boltzmann equation force term: 

\begin{equation}
	{\mathbf a}\cdot\nabla_\xi n 
	 \label{FORCE}	
\end{equation}

\noindent and the equilibrium distribution in Hermite polynomials in
the lattice velocities. Then they rearranged the acceleration
${\mathbf a}$ such that it explicitly modifies the macroscopic
velocity in the equilibrium distribution, leaving a term linear in
${\mathbf a}$. If only linear terms were to appear in
Eq. (\ref{SHAN-CHEN-EQDIST}), the Shan-Chen prescription for an
interparticle force would then coincide with the way it is included in
the continuum BGK-Boltzmann equation, as pointed out by Luo and Martys
{\em et al.}. To this end, following Nekovee {\em et al.}
\cite{MAZIAR-ET-AL}, we simply drop from Eq. (\ref{SHAN-CHEN-EQDIST})
any term nonlinear in the acceleration ${\mathbf a}$. We thus obtain a 
modified Shan-Chen collision term, which is why our model is
termed modified Shan-Chen. The modified Shan-Chen collision term is:

\begin{equation}
	\Omega_k^{\prime\alpha} 
	\equiv
	\Omega_k^\alpha	 +
	\sum_{\bar\alpha}\sum_l  \Lambda_{kl}^{\alpha\bar\alpha}
	n_l^{\bar\alpha}				\,,
\label{LBEQUATION}
\end{equation}

\noindent where

\begin{equation}
	\Lambda_{kl}^{\alpha\bar\alpha} = \omega_k \Big[
	\frac{1}{c_s^2}(\delta_{\alpha\bar\alpha}{\mathbf
	c}_k-\zeta_{\alpha\bar\alpha}{\mathbf c}_l)  +
	\zeta_{\alpha\bar\alpha} \frac{{\mathbf c}_k\cdot{\mathbf
	c}_l}{c_s^4}{\mathbf c}_k  \Big]
	\cdot{\mathbf a}^\alpha\,\tau^\alpha
	\label{LAMBDA}
\end{equation}

\noindent and

\begin{equation}
	\zeta_{\alpha\bar\alpha} = \frac{n^\alpha}{n^{\bar\alpha}} \,
	{\frac{\rho^{\bar\alpha}}{\tau^{\bar\alpha}}}/
	{\sum_{\bar\alpha}
	\frac{\rho^{\bar\alpha}}{\tau^{\bar\alpha}}}	\,,
\end{equation}

\noindent where we have made use of the condition
(\ref{CREEPING_FLOW}) below. The second term arising in
(\ref{LBEQUATION}) accounts for interparticle interactions other
than the binary collisions implicit  in the Boltzmann collision term,
$\Omega$ \cite{BGL}. This includes a collision operator
$\Lambda_{kl}^{\alpha\bar\alpha}$ resulting from mean-field
interactions among different fluid components \cite{MAZIAR-ET-AL},
which gives rise to phase separation for immiscible multicomponent
systems.

The inclusion of a mean-field force in the Shan-Chen model leads to
the breakdown of the local momentum conservation that holds for
noninteracting ideal gases, {\em cf.} Subsection
(\ref{IDEAL_GASES}). However, the forces felt by neighbouring portions
of fluid follow an action-reaction mechanism that leads to global
momentum conservation (i.e. over the whole lattice). This was
numerically confirmed for our third-order-equilibrium, modified
Shan-Chen model too. 

It can be shown that the condition for momentum conservation in the
absence of interactions, Eq. (\ref{FLUID_VELOCITY}), leads to that
needed when using a second-order expansion of the equilibrium
distribution, namely

\begin{equation}
	{\mathbf u} = {\mathbf v}			\,,
	\label{FLUID_VELOCITY_2ND-ORDER}
\end{equation}

\noindent only in the limit of creeping flows to second order, i.e.

\begin{equation}
	u^2 \approx 0					\,.
	\label{CREEPING_FLOW}
\end{equation}

\noindent We therefore implemented the computation of the velocity
according to (\ref{FLUID_VELOCITY_2ND-ORDER}) rather than
(\ref{FLUID_VELOCITY}). The condition (\ref{CREEPING_FLOW}) is
satisfied, as global momentum would not be conserved otherwise. In
addition, we confirmed in our simulations that the fluid velocity was 
kept under 28\% of the speed of sound by 67\% of the lattice
nodes. This means squared Mach numbers under 0.08. This purports to
show that the expansion to third order, implemented in this  model to
extend the parameter space for which the equilibrium distribution
remains positive, for momentum conservation at least adds   very
little.

In our LB model, the kinematic viscosity of the mixture is given by 

\begin{equation}
	\nu = \frac{\eta}{\rho} = c_s^{-2}\Big(\sum _\alpha x
	_\alpha\tau _\alpha -\frac{1}{2}\Big) \label{nu}
\end{equation}

\noindent where $c_s^{-2} =3$ for our lattice,  $\tau _\alpha$ is the
relaxation time of the $\alpha$th component and $x _\alpha$ is its
mass concentration defined as $\rho _\alpha/\rho$
\cite{SHAN-CHEN}. For a region of pure $\alpha$th component,

\begin{equation}
\nu = \frac{1}{3}\Big(\tau -\frac{1}{2}\Big)
\label{KINEM-VISCOSITY}
\end{equation}

\noindent which also holds for our case of a 50:50 mixture, for which

\begin{equation}
	\sum_\alpha x_\alpha\tau^\alpha =  \tau\sum_\alpha x_\alpha =
	\tau
\end{equation}

\noindent since all relaxation times are the same.

\section{The surface tension}
\label{SURFTENS}

The surface tension $\sigma$ arises as an
emergent effect due to intercomponent interactions. It is calculated
by measuring the components of the pressure tensor ${\mathsf
P}=\{P_{ij}\}$ across a planar interface perpendicular to the $z$-axis
through the formula 

\begin{equation}
	\sigma = \int_{-\infty}^{+\infty} \Big[ P_{zz}(z) - P_{xx}(z)
	\Big] \mathrm{d}z
\end{equation}

\noindent where $P_{ij}$ is the flux of the $i$th component of the
momentum across a surface perpendicular to the $j$th cartesian
axis. This pressure tensor, consistent with the force
(\ref{LBE-FORCE}), is

\begin{eqnarray}
	{\mathsf P}({\mathbf x})
	&=&
	\sum_{\alpha}\sum_k\rho_k^{\alpha}({\mathbf x}) {\mathbf c}_k
	{\mathbf c}_k 				\\ 	\nonumber
	&+&
	\frac{1}{4}\sum_{\alpha,\bar\alpha} g_{\alpha\bar\alpha}
	\sum_{{\mathbf x}'} \Big[ \psi^{\alpha}({\mathbf x})
	\psi^{\bar\alpha}({\mathbf x}') + \psi^{\bar\alpha}({\mathbf
	x}) \psi^{\alpha}({\mathbf x}') \Big]
	(\mathbf{x-x'})(\mathbf{x-x'})				\,, 
	\label{PRESSURETENSOR}
\end{eqnarray}

\noindent with ${\mathbf x}^\prime\equiv{\mathbf x}+{\mathbf c}_k$ in
this study. This leads to the same expression for the scalar pressure
as that in the momentum balance equation obtained by multiplying the
LB equation (\ref{BGK-OP-NOFORCE}) using the collision term
(\ref{LBEQUATION}) by ${\mathbf c}_k$ and summing over $k$. Here,  
$\rho_{k}^{\alpha}({\mathbf x})$ is the mass density of species
$\alpha$  with velocity ${\mathbf c}_k$ at the site ${\mathbf
x}$. Eq. (34) contains a {\it kinetic term} due to the
free streaming of particles corresponding to an ideal gas
contribution, plus a {\it potential} or {\it virial term} due to the
momentum transfer among particles of equal and distinct colour,
through the interparticle force \cite{FERZIGER}.

As previously noted, the surface tension in the modified Shan-Chen
model is an emergent, hence not directly tunable quantity, in
contradistinction to the situation with free-energy based
lattice-Boltzmann models. It depends on the density $\rho$, the
coupling $g$ and the relaxation time $\tau$, and has to be determined
by simulation. We computed its dependence on these parameters to be as
follows 

\begin{equation}
	\frac{\partial\sigma}{\partial\rho} > 0 	\,,\qquad
	\frac{\partial\sigma}{\partial g} > 0 		\,,\qquad
	\frac{\partial\sigma}{\partial\tau} < 0 	\,.
	\label{SIGMA-STEERING}
\end{equation}

\noindent This behaviour is useful when steering through the parameter
space in search of specific values of $L_0$ and $T_0$. Numerical
results on the surface tension are reported in the next Section.

\section{Simulations}
\label{SIMUL}

We restrict ourselves to critical (50:50) mixtures, which are the type
of configurations leading to a spinodal decomposition process as
opposed to nucleation. Experimentally, spinodal decomposition is
characterised by long-wavelength, infinitesimal density perturbations
which are unstable after the quench, hence favouring the segregation,
whereas nucleation generally presents short wavelength, finite
perturbations, and metastability is not uncommon. Nucleation is hence
a more complex phenomenon which is usually considered after an initial
study in spinodal decomposition has been performed. 
   
We aim at reproducing the early time diffusive and later time viscous
and inertial regimes predicted by carrying out scaling analyses on the
Cahn-Hilliard Model-H equations \cite{GUNTON,BRAY}. Growth laws
predicted for those are $l\propto t^{1/3}$, $l\propto t$ and $l\propto
t^{2/3}$,  respectively. Under the assumptions of the dynamical scaling
hypothesis made in the introduction, those regimes are uniquely
characterised by the length and time

\begin{equation}
	L_0=\frac{\rho}{9\sigma(\rho,\tau,g)}
	\Big(\tau-\frac{1}{2}\Big)^2\,, \qquad
	T_0=\frac{\rho^2}{27\Big(\sigma(\rho,\tau,g)\Big)^2}
	\Big(\tau-\frac{1}{2}\Big)^3\,,
\end{equation}

\noindent obtained by inserting (\ref{KINEM-VISCOSITY}) into
(\ref{L0T0}).

Having in mind keeping simulation time at a minimum, the values of
$\rho$, $\tau$ and $g$ must be such as to allow the fluids to be
immiscible and approach equilibrium quickly whilst ensuring numerical
stability and positive shear viscosity. This amounts to keeping $\rho$
as high as possible, $\tau$ close to $1/2$, 
and $g$ as large as allowed by the onset of numerical instabilities
which set in when the forcing term is too large. A large $g$ allows
for the early time transient, dominated by diffusion, to be of short 
duration. Finally, seeking the diffusive regime means looking at
very early times, which is attained for large values of
$T_0$. Conversely, the hydrodynamic inertial behaviour requires as
small values of $T_0$ as possible.

In Table \ref{PARAMETERS} we present the parameters selected in this
study, along with the measured surface tension. We also include the
length and time scales associated with them, which are used to compute
dimensionless lengths and times in the model.

\begin{table}[!htb]
\begin{center}
\begin{tabular}{|c|c|c|c|c|c|c|}
\hline  Parameter set 	& $\rho$ & $\tau$ & $g$
& $\sigma(\rho,\tau,g)$	& $L_0(\rho,\tau,g)$ & $T_0(\rho,\tau,g)$\\
\hline  
{\tt I   }  		&  0.8   &2.000   & 0.06 
& 0.002059	& 97.1    &  18870 	\\ 
{\tt II  } 		&  0.8   & 1.500  & 0.06 
& 0.004777   	& 18.6   &  1038.8 	\\  
{\tt III }		&  0.8   & 1.000  & 0.06 
& 0.010292   	& 2.16    &  28.0    	\\     
{\tt IV  }  		&  0.8   & 0.625  & 0.05 
& 0.017458	& 0.0796  &  0.152   	\\ 
\hline
\end{tabular}
\end{center}
\caption{\small Model parameters studied, including the surface
tension $\sigma(\rho,\tau,g)$ measured for a planar interface on a
$4\times 4\times 128$ lattice, and the characteristic length $L_0$ and 
time $T_0$ for each parameter set. The existence of the latter two is
based on the validity of the dynamical scaling hypothesis, and that
diffusive currents are negligible with respect to hydrodynamic
currents and capillary forces.}   
\label{PARAMETERS}
\end{table}

The initial condition used for all the simulations was a thorough
mixture of the two phases, with randomly distributed fluctuations. To
realise this, each velocity direction $k$ at each lattice site was
populated with one density $\rho_k^\alpha({\mathbf x},t)\equiv
m^\alpha n_k^\alpha({\mathbf x},t)$ for each species 
$\alpha=\mathrm{R},\mathrm{B}$  as a white-noise, pseudo-random
floating point number between 0.0 and 0.8, where $m^\alpha$ are the
particle masses, all set to unity. Note that the density $\rho$ in
Table \ref{PARAMETERS} is defined as the lattice average 

\begin{equation}
	\rho
	\equiv
	\langle\rho^\mathrm{R}({\mathbf x},t)
	+
	\rho^\mathrm{B}({\mathbf x},t)\rangle		\,,
\end{equation}

\noindent where $\rho^\alpha({\mathbf
x},t)\equiv\sum_k\rho_k^\alpha({\mathbf x},t)$, and due to the
critical composition we use, amounts to the maximum value of either of
the summands. 

Lattice sizes used were $128^3$ and then $256^3$ 
to check for finite size effects. Simulations for $128^3$ systems were
run for 700 or 1400 time steps, and for 200 or 250 time steps for
$256^3$ systems, depending on the parameter set. Following Kendon {\em
et al.}'s prescription to keep finite size effects at bay
\cite{KENDON99}, we neglected domain sizes larger than a quarter of the
lattice side size. There is no reason {\it a priori} to choose this
particular threshold. As we shall see, this allows the generation of a
domain size range broad enough for data acquisition; furthermore,
finite size effects were quantified by using the two aforementioned
lattice sizes. 

Surface tension was measured on $4\times 4\times 128$ and $16\times
16\times 128$ lattices, allowing plenty of room along the
non-isotropic direction $z$ for the fluid's physical quantitites to
achieve values characteristic of the bulk before being affected by the
second interface with periodic boundary conditions imposed. We found
that the surface tension did not vary by more than 1\% when the length
along the $z$ direction was doubled, which is the only direction where
we would expect any variation as translational symmetry is broken.

To compute the average domain size, Eq. (\ref{DOMAIN-SIZE}), we
perform discrete Fourier transforms. The sampling theorem \cite{NR}
warns us to ensure that our fluid mixture does not exhibit spatial
frequencies larger than the Nyquist critical frequency $f_c$, defined
as half the sampling frequency. This not being the case, the power
spectrum in the interval $[0,f_c]$ is altered by frequencies larger
than $f_c$ as a result of aliasing. Because the sampling frequency on
the lattice is one, the maximum frequency any relevant quantity of our
fluid mixture is allowed to  have according to the sampling theorem is
$1/2$, i.e., of  wavelength two. This means that any spatial variation
is bound  to happen between two contiguous lattice sites, which is
something we already knew: the resolution of the lattice is finite and
dictated by the lattice size. We used the FFT routine {\tt rlft3()}
for real, 3D data sets \cite{NR}.

Calculation of the reduced time $t$ requires an assessment of
$T_{\mathrm{int}}$. $T_{\mathrm{int}}$ serves to redefine the
time such that the domains have zero size at the time origin, which is
not the case in the actual simulations. Depending on the regime
reached by the parameter set employed, domains may start to grow
immediately after time step zero, completely avoiding the diffusive
stage.

We assess $T_{\mathrm{int}}$ in the following way. We first compute
the intersection with the abscissae of a linear fit interpolating all
data starting after the initial purely diffusive transient is
completed, that is, for which interfaces are thin enough and $L(T)$ just
starts to grow. The intercept is used as an initial guess for $a_1$ in
a Levenberg-Marquardt non-linear least-squares fit of the form

\begin{equation}
	y=a_0(x-a_1)^{a_2}.  \label{NL-FIT}
\end{equation}

Once $T_{\mathrm{int}}$ is computed, and the data sets are normalised
by $L_0$ and $T_0$, hence becoming $(t,l)$ data pairs, we perform fits
to the function (\ref{NL-FIT}) to determine the growth exponent
$a_2$. Initial guesses for the fitting coefficients are $a_0^0=1.0$,
$a_1^0=0.1$, and $a_2^0=1.0$. The tolerance for these fits was set to
$10^{-5}$, this being defined as the unsigned increment of $\chi^2$
between two consecutive iterations, divided by the number of degrees
of freedom.

Uncertainties in parameters are also taken care of. Because standard
errors $\{\Delta_k S\}$ are incurred in the structure function
spherical averaging (\ref{SASF}), these transmit down to $L$ and $l$,
and to $t$ through the determination of $T_{\rm{int}}$. In this study,
however, errors in the abcissae are disregarded as they do not depend
on time, and therefore represent equal weights for data points in the
least-squares functional to minimise.

We performed the simulations using a number of processors ranging from
32 to 128 on a Cray T3E-1200E and on SGI Origin2000 and Origin3800
supercomputers. The code  is an implementation in Fortran90 using the
Message Passing Interface (MPI) as parallelisation protocol, and it
shows scaling with the number of processors between 50\% to 90\% of
linearity on the Cray T3E platform \cite{NEKOVEE_MEXICO} up to 64 PEs,
and better behaviour on SGI Origin platforms. CPU times used up to run
a $128^3$ lattice for 1400 time steps, or a $256^3$ lattice for 250
time steps, ranged between 3 to 6 hours per parallel process.

An important issue in dealing with the lattice sizes employed here is
to have access to massive disk storage. For our largest lattices,
1.9 Gbytes of measurements were dumped onto disk at each measurement
time step. A lattice of $256^3$ sites run for 700 time steps,
measuring every 25, requires 40 Gbytes to store the order parameter, the
density fields for each phase, momenta, and checkpoint files, the
latter being needed if we wish to restart the simulation at the point where
it stops. To that we need to add some additional working space for
converting the dumped binary data into machine-portable XDR
format \cite{XDR}. For this work we required 200 Gbytes on disk, plus tape
storage to free up space when required. XDR files were visualised
using the commercial package AVS \cite{AVS}. 

It is worth noting that our results did not undergo a process of
lattice size reduction, in the sense of averaging over
nearest-neighbouring sites in order to deal with limited computational
resources, as was done in previous studies on 3D spinodal
decomposition \cite{KENDON01,CATES_PRIVATE}. Hence, we benefitted from
measuring and visualising all data output from our
simulations. Current limitations in computing resources  
prevented us from simulating lattices of $512^3$ or $1024^3$ sizes,
which would otherwise be desirable in order to decrease the fluid's
minimum Knudsen number, helpful in reaching for the thermohydrodynamic
limit as a multiscale Chapman-Enskog expansion procedure
shows. However, this situation is bound to change soon with the advent  
of terascale computing capabilities.

\subsection{Growth exponents}

%% Parameter sets I, II, III and IV

\begin{figure}[p]
\begin{center}
\includegraphics[angle=-90,width=16cm]{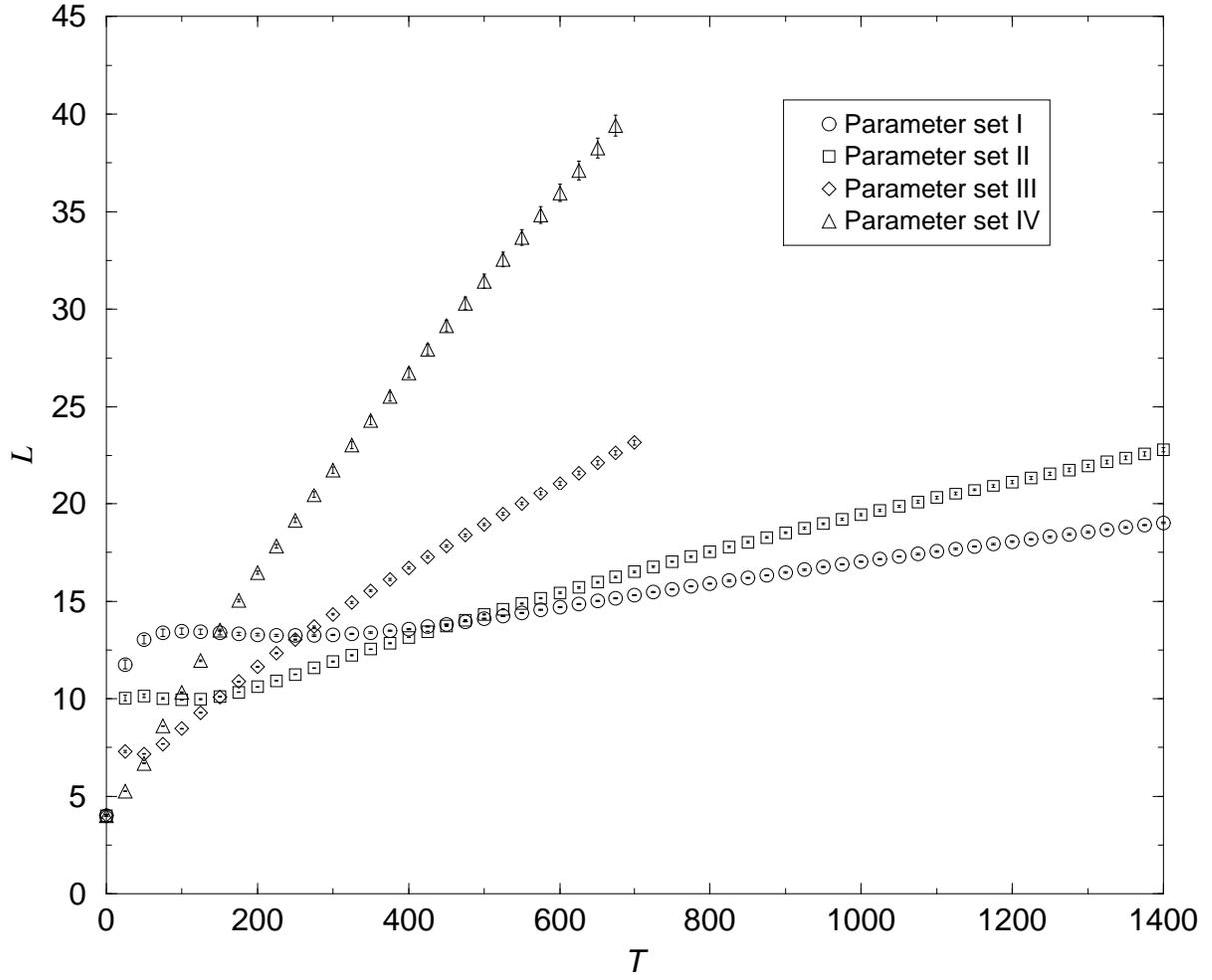}
\caption{Evolution of the average domain size for parameter sets I,
II, III, and IV ({\em cf.} Table \ref{PARAMETERS}) with the time
step. Error bars are included and represent the uncertainty
transmitted from the standard error of the structure function
spherical average. Lattice size is $128^3$. All quantities are reported
in lattice units.}     
\label{SIZ}
\end{center}
\end{figure}

\begin{figure}[p]
\begin{center}
\includegraphics[angle=0,width=16cm]{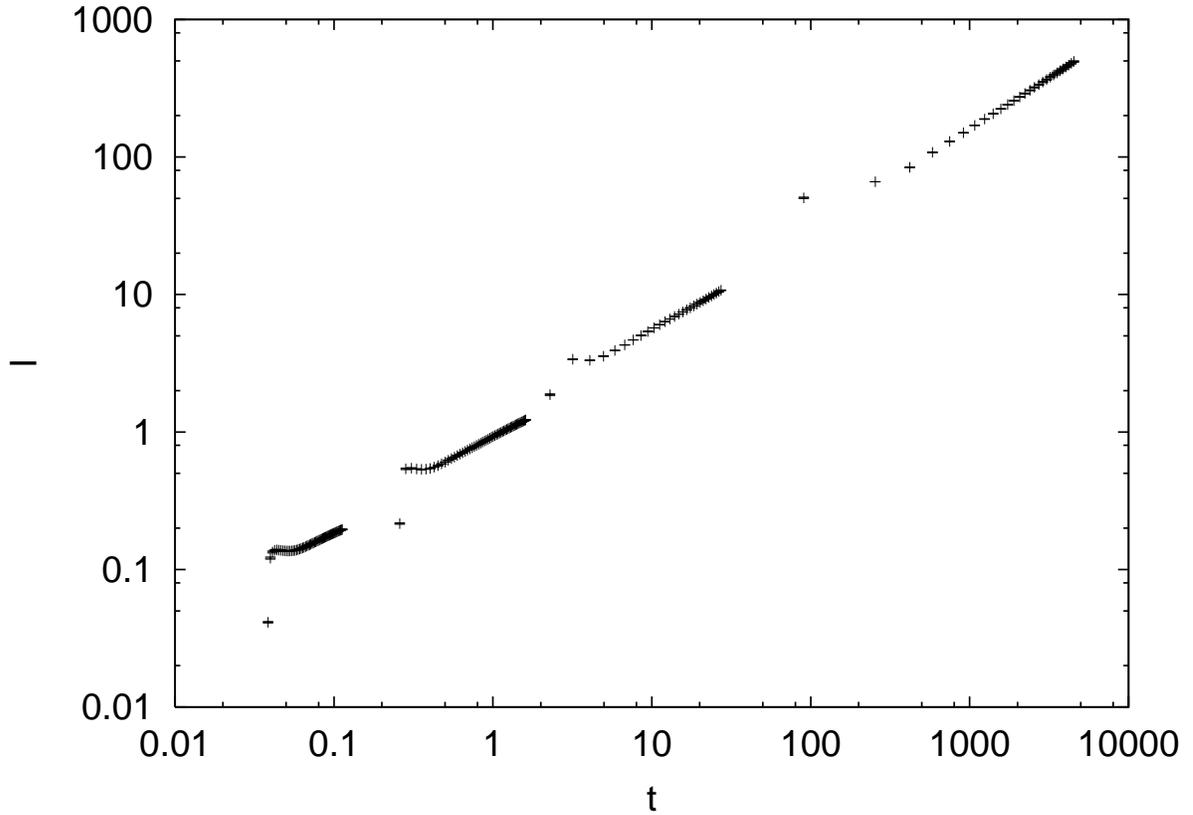}
\caption{Log-log plot of reduced length versus reduced time for the
$128^3$-lattice data sets. Error bars are included. The four data sets 
correspond to parameter sets I, II, III, and IV ({\em cf.} Table
\ref{PARAMETERS}), from left to right.  Viewed from a grazing angle,
one can see that a simple, algebraic interpolating curve is not truly
obtainable here. The first few points of each
set correspond to diffusive, zero-growth stages. The units on both
axes are dimensionless.}  
\label{ALL_REGIMES128}
\end{center}
\end{figure}

Figure \ref{SIZ} shows the average domain size in lattice units as
obtained straight from the simulations, for all parameter sets ({\em
cf.} Table \ref{PARAMETERS}). Reynolds numbers achieved for each of
these are $\mathrm{Re}=0.18$, $0.49$, $2.7$, and $37$. For parameter
set I, we can see that after a transient during which there is a rapid
mass convection to nearest neighbours, domain growth flattens out and
starts growing at about $T=400$. We will look at this in further
detail; for now it can be seen that the breadth of the plateau
decreases with the Reynolds number. Finally in
Fig. \ref{ALL_REGIMES128} we show the same curves after rescaled to
$L_0$ and $T_0$, in reduced units. 

Fits to the model $y=a_0(x-a_1)^{a_2}$ for Fig. \ref{ALL_REGIMES128}
are given in Table \ref{TABLE_EXPONENTS}, and they proved to be quite 
sensitive to the number of points fitted. Domain growth shows an
increasing segregation speed, $t^{0.545}$, $t^{0.593}$, $t^{0.623}$,
and $t^{0.717}$, with increasing Reynolds number. These data
sets correspond to characteristic lengths and times in the ranges
$0.0796 < L_0 < 97.1$ and $0.152 < T_0 < 18870$. These contain 
the values for which Kendon {\it et al.} \cite{KENDON99} observed a
viscous linear exponent, $L_0=5.9$ and $T_0=71$. This, therefore,
invalidates the universality of the dynamical scaling hypothesis. 

By looking at Fig. \ref{ALL_REGIMES128} from a grazing angle one can
easily see that a simple, algebraic interpolating curve is not
obtainable here. Kendon {\em et al.} \cite{KENDON99,KENDON01} and
Pagonabarraga {\em et al.} \cite{PAGONA01,PAGONA-PRIVATE} used a
method to improve this curve. They left $T_{\mathrm{int}}$ as an
adjustable fitting parameter such that there is a reasonable collapse
onto a simple, single algebraic curve for all parameter sets
simulated; from this they obtained a window of $T_{\mathrm{int}}$ in
which collapse is reasonable. Then they checked whether the different
values for $T_{\mathrm{int}}$ from each individual parameter set lay
within such a window. Quoting Kendon {\em et al.} ({\em cf.}
Section 9.3 in Ref. \cite{KENDON01}), ``although this [procedure] is
capable of falsifying the scaling hypothesis [...], its
non-falsification [...] may not represent persuasive proof that  the
scaling is true.'' We adhere to this comment and prefer not to
manipulate the data sets in such a way.

\begin{table}[!htb]
\begin{center}
\begin{tabular}{|c|r @{$\pm$} l|r @{$\pm$} l|r @{$\pm$} l|c|}
\hline
Parameter set 			& 
\multicolumn{2}{c|}{$a_0$} 	& 
\multicolumn{2}{c|}{$a_1$} 	& 
\multicolumn{2}{c|}{$a_2$} 	& 
$\chi^2/\mathrm{ndf}$  			\\ 
\hline
{\tt I  } & 0.644&0.014 
	  & $-2\times10^{-5}$&0.002
	  & 0.545&0.014
	  & 0.46			\\
\hline
{\tt II } & 0.924&0.004 
	  & $6\times10^{-6}$&0.007
	  & 0.607&0.006
	  & 1.23			\\
	  & 0.922&0.003 
	  & $-2\times10^{-5}$&0.007
	  & 0.593&0.007
	  & 0.48			\\ 
\hline
{\tt III} & 1.248&0.031 
	  & $-0.007$&0.100
	  & 0.650&0.007
	  & 2.71			\\
	  & 1.362&0.010 
	  & $-1\times10^{-4}$&0.03
	  & 0.623&0.002
	  & 0.68			\\
\hline
{\tt IV } & 0.941&0.019 
	  & 0.01&3.9
	  & 0.743&0.002
	  & 0.10			\\
	  & 1.139&0.017 
	  & $-0.01$&3.6
	  & 0.717&0.002
	  & 0.14			\\
\hline
\end{tabular}
\end{center}
\caption{\small Levenberg-Marquardt nonlinear least-squares fits of
$\,l\,\,\mathrm{vs}\,\,t\,$ data to model (\ref{NL-FIT}), for each
parameter set attempted. The first line for each set belongs to $128^3$
data, the second line to $256^3$ data, the latter being unavailable
for set I. Fitting parameters are given, plus the weighted sum of
squared residuals ($\chi^2$) divided by the fit's number of degrees of
freedom. Weights are the inverses of squared uncertainties. Note that
$\chi^2/\mathrm{ndf}$, also referred to as the variance of residuals,
is expected to approach unity. Values larger than $1.0$ may be due to
an insufficient number of data points, data errors not normally
distributed, or an incorrect model function. Values smaller than $1.0$
may be the result of too large error bars, or too general a model
function.}   
\label{TABLE_EXPONENTS}
\end{table}

%%%%%%%%%%%%%%%%%%%%%%%%%%%%%%%%%%%%%%%%%%%%%%%%%%%%%%%%%%
%% P L O T S     A N D     P I C T U R E S
%%%%%%%%%%%%%%%%%%%%%%%%%%%%%%%%%%%%%%%%%%%%%%%%%%%%%%%%%%

%%% GROWTH EXPONENTS: 

%%%%%%% .SIZ PLOTS

%%% GROWTH EXPONENTS: ALL REGIMES

%%%%%%% 128

\subsection{Structure function}

%% PARAMETER SET I

\begin{figure}[p]
\begin{center}
\includegraphics[angle=-90,width=16cm]{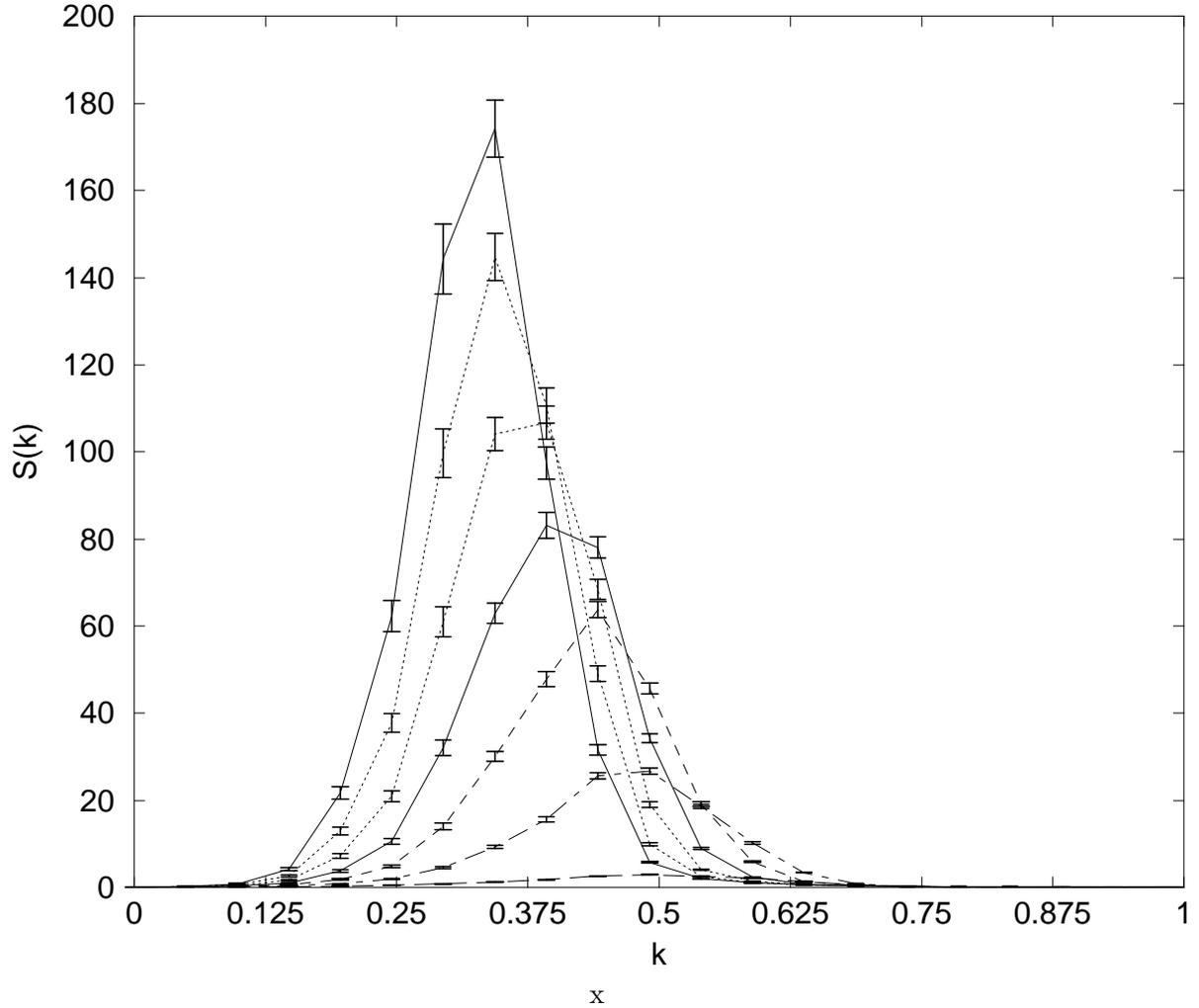}
x\caption{Spherically averaged structure function versus wavenumber,
for parameter set I ({\em cf.} Table \ref{PARAMETERS}). $128^3$
lattice. Error bars represent the standard error of the structure
function spherical average. Time slices shown are time step 200, 400,
600, 800, 1000, 1200, and 1400 from right to left. All quantities are
reported in lattice units.}  
\label{DIFF128_SF_vs_k}
\end{center}
\end{figure}

\begin{figure}[p]
\begin{center}
\includegraphics[angle=-90,width=16cm]{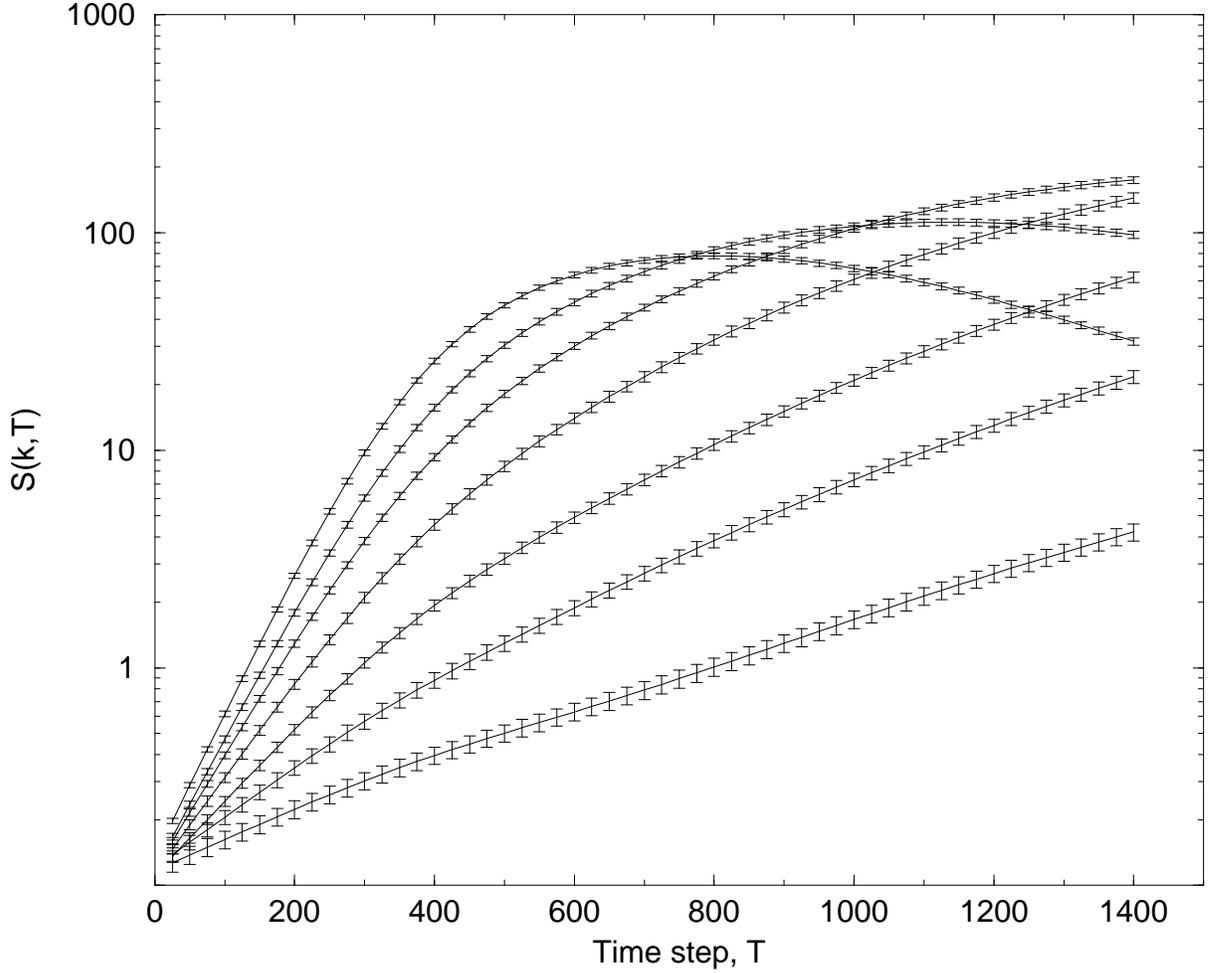}
\caption{Evolution of the spherically averaged structure function with
the time step for parameter set I and a $128^3$ lattice, on a
logarithmic scale. When observed along the ordinate $T=200$, the curves
correspond to wavenumbers $k=0.147$, 0.196, 0.245, 0.295, 0.344,
0.393, and 0.442 from bottom to top, respectively. Error bars
represent the standard error of the structure function spherical
average. Regions of linear growth are those for which the exponential
behaviour Eq.(\ref{EXP_GROWTH}) holds. For wavenumbers up to 0.2
exponential and therefore diffusive behaviour is seen for all the
simulation time. For larger wavenumbers (and hence smaller domain
sizes) diffusion occurs as a transient. All quantities are reported in
lattice units.}      
\label{DIFF128_SF_vs_T_k0.15--0.45}
\end{center}
\end{figure}

\begin{figure}[p]
\begin{center}
\includegraphics[angle=-90,width=16cm]{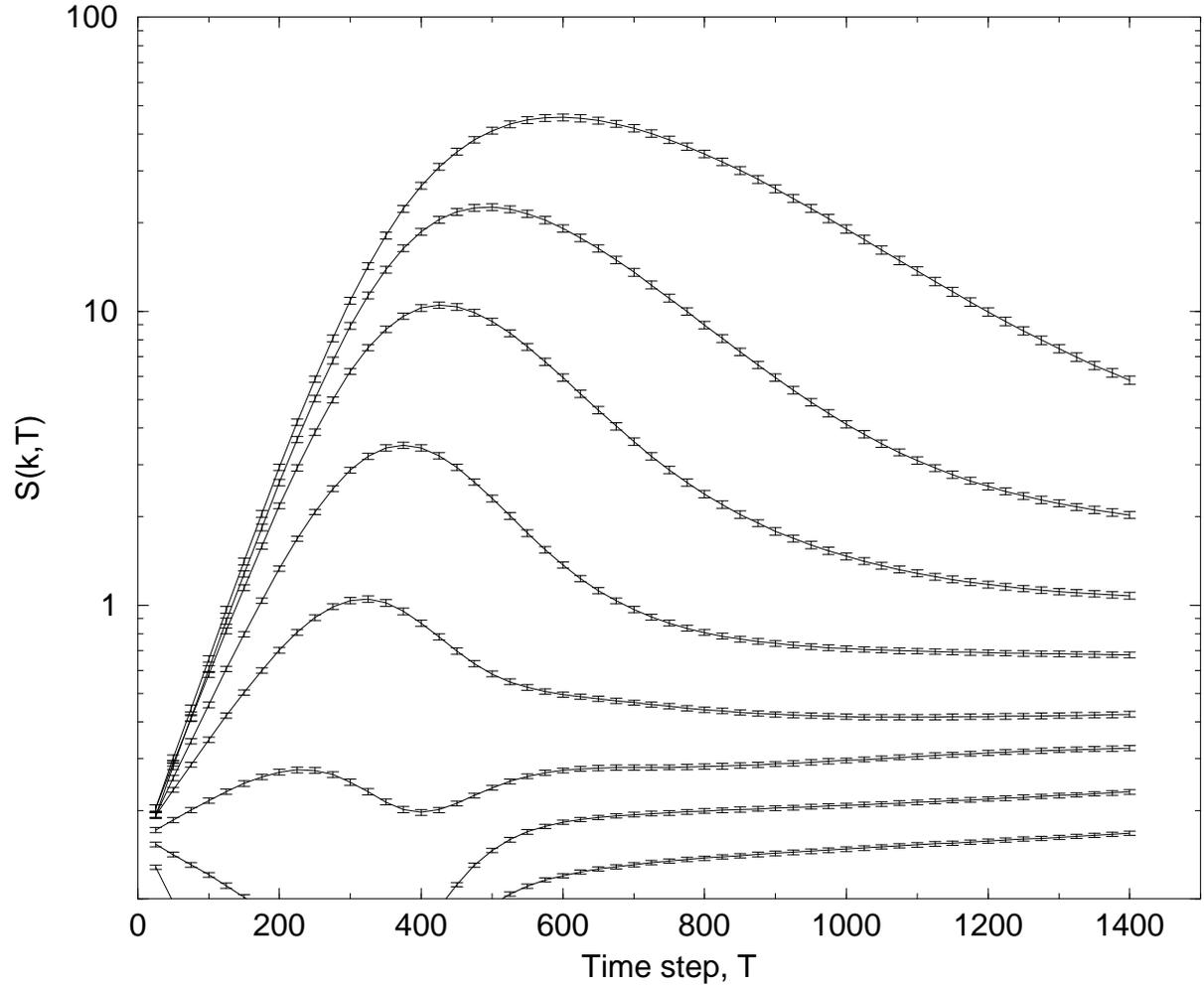}
\caption{Similar to Fig \ref{DIFF128_SF_vs_T_k0.15--0.45}, but the
curves correspond to wavenumbers $k=0.491$, 0.540, 0.589, 0.638,
0.687, 0.736, 0.785, and 0.834 from top to bottom, respectively. We
can see that linear growth ceases to hold for wavenumbers larger than
about 0.736, in accordance with existence of an upper cutoff for the
validity of Eq. (\ref{EXP_GROWTH}). All quantities are reported in
lattice units.}  
\label{DIFF128_SF_vs_T_k0.45--0.75}
\end{center}
\end{figure}

For parameter set I ({\em cf.} Table \ref{PARAMETERS}) we show in
Fig. \ref{DIFF128_SF_vs_k} a family of spherically-averaged structure
functions versus wavenumbers, corresponding to time steps 200, 400,
600, 800, 1000, 1200, and 1400, from right to left. Just as in
scattering cross-section measurements \cite{GUNTON}, we observe the
peaks to grow and approach  small wavenumbers as time evolves. In 
Figs. \ref{DIFF128_SF_vs_T_k0.15--0.45} and
\ref{DIFF128_SF_vs_T_k0.45--0.75} we show the same family of curves
using time steps as absissae and wavenumbers as parameters. Regions of
linear growth with time on such a logarithmic scale indicate that a
diffusive process is dominating the dynamics. In fact, an exponential
time growth for the structure function shortly after the quench below
the spinodal curve was predicted from the linearised Cahn-Hilliard
Model-B equations without noise \cite{GUNTON}, which although
incorporating order-parameter conservation, does not include
hydrodynamics.  This Cahn-Hilliard equation
might be applicable to regimes in our fluid where hydrodynamic effects
were unimportant, as in the initial stages. Assuming linear
perturbations $\phi^\prime$ to the order parameter, Cahn predicted
that for fluctuations of small amplitude and long wavelength there is
an instability of the shape 

\begin{equation}
	S(k,t) = S(k,0)\mathrm{e}^{-2\omega(k)t} \label{EXP_GROWTH}
\end{equation}

\noindent for $k<k_c$, where $k_c$ depends on the diffusion
constant. Here, $t$ is the time, $\omega(k)<0$, and
$S(k,t)\propto\langle\Big|\phi_{\mathbf k}^\prime(t)\Big|^2\rangle$,
the brackets denoting averaging in reciprocal space over a shell of
radius $k$.

Exponential growth occurs in our simulations as can be seen from
Figs. \ref{DIFF128_SF_vs_T_k0.15--0.45} and
\ref{DIFF128_SF_vs_T_k0.45--0.75} for about the first 350 time steps
for most of the wavenumbers, indicating its transient character. The
plateau of Fig. \ref{SIZ}, Set I, lasts during the first 400 time
steps, and we can see, Fig. \ref{DIFF128_SF_vs_k}, that up to 400 time
steps the peak in the structure factor varies in height and very
little in wavenumber, and is located at 0.491 (lattice units). This
leads us to think that at these early stages the dynamics is mainly
making walls thinner while average domain sizes barely change. In
addition, visual inspection of the order parameter confirms the
latter and suggests that hydrodynamic currents are weak, leaving
diffusion as the mechanism leading the phase segregation process. When
we check the structure function temporal evolution,
Figs. \ref{DIFF128_SF_vs_T_k0.15--0.45} and 
\ref{DIFF128_SF_vs_T_k0.45--0.75}, for the curves at and around
$k=0.491$, we see that up to exactly 400 time steps do they show
exponential growth, as the Cahn-Hilliard Model B predicts for a
diffusive scenario. Also, exponential growth does not hold for all
wavenumbers, but only for those smaller than about 0.7, in agreement
with the existence of an upper cutoff for the validity of
Eq.(\ref{EXP_GROWTH}), predicted from Model B.

However, not all the wavenumbers follow Model B's predictions, namely,
that exponential growth is a transient and occurs up to a threshold
wavenumber. In fact, exponential growth holds for all the 
time steps of the simulation for the larger length scales (wavenumbers
up to about 0.245), suggesting that diffusion never ceases to
dominate their dynamics. Also, exponential growth is seen for very
small domain sizes (wavenumbers larger than 0.736) for time steps well 
advanced in the coarsening dynamics, after 600 time steps. These
wavenumbers are close to and above the expected Model-B upper cutoff
for exponential growth, set by the change in slope from positive to
negative in Fig. \ref{DIFF128_SF_vs_T_k0.45--0.75}. These departures
from Model B's predictions hold nonetheless for domain sizes far from
the first moment of the structure factor, which is close to its peak
and is our average domain size measure. It would be desirable in
future works to investigate diffusional processes at $k<0.245$ for all
of the simulation time, and $k>0.736$ at late times: according to the  
Cahn-Hilliard linearised Model B, for these cases diffusion is
negligible or forbidden, respectively.

%% PARAMETER SET II, III,  IV

\begin{figure}[p]
\begin{center}
\includegraphics[angle=-90,width=16cm]{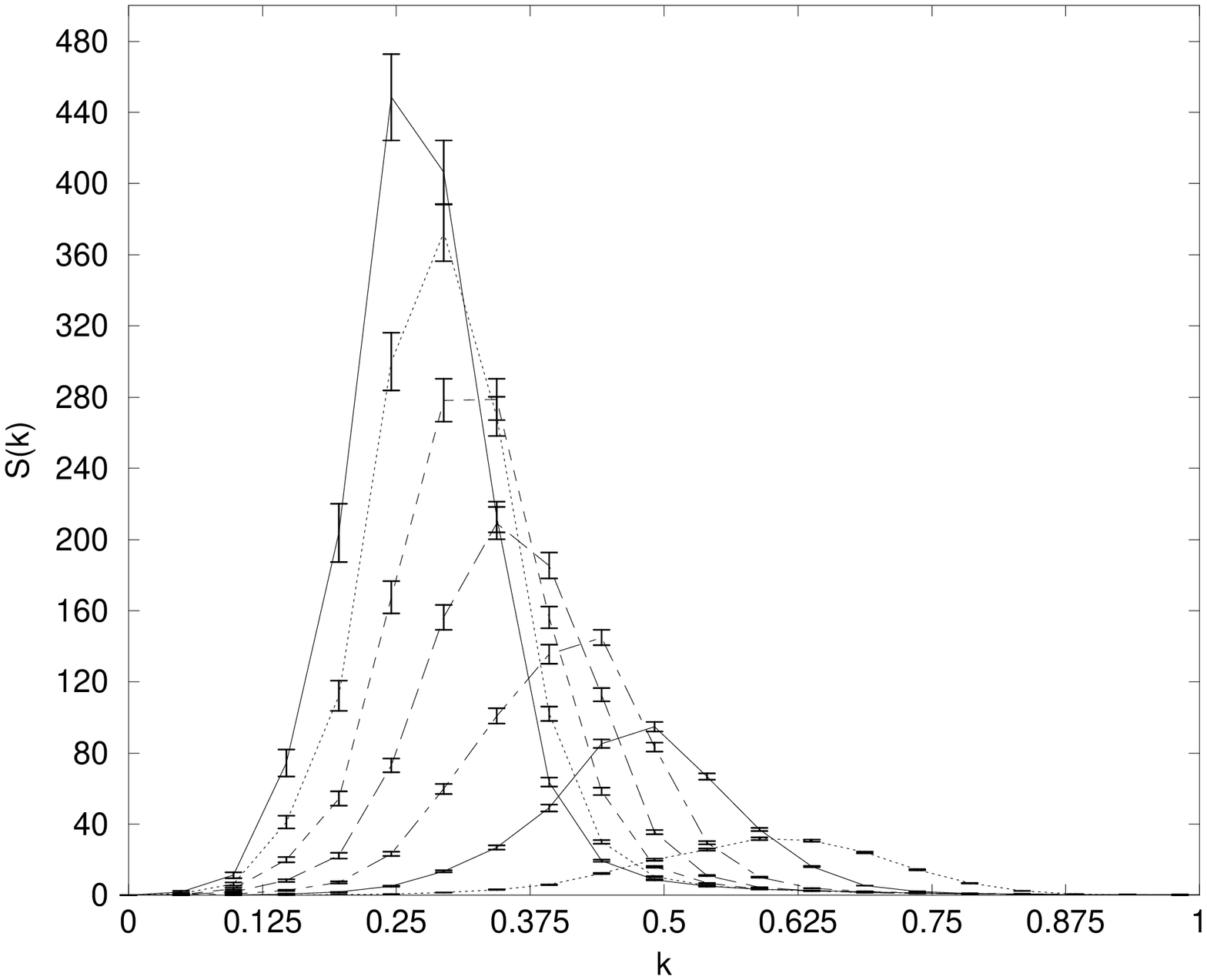}
\caption{Spherically averaged structure function versus wavenumber,
for parameter set II ({\em cf.} Table \ref{PARAMETERS}). $128^3$
lattice. Error bars represent the standard error of the structure
function spherical average. Time slices shown are time step 200, 400,
600, 800, 1000, 1200, and 1400 from right to left. All quantities are
reported in lattice units.}  
\label{DIFFB128_SF_vs_k}
\end{center}
\end{figure}

\begin{figure}[p]
\begin{center}
\includegraphics[angle=-90,width=16cm]{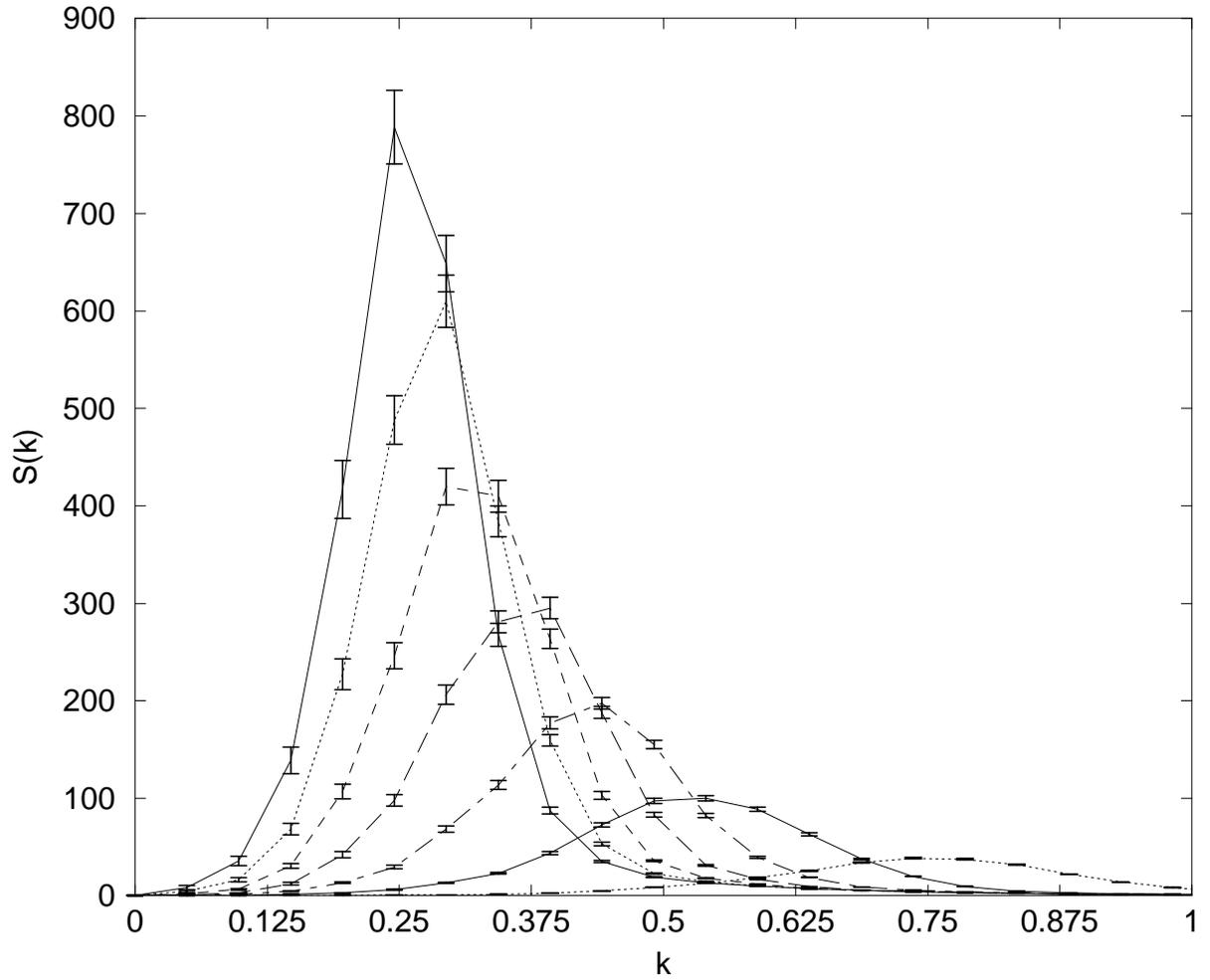}
\caption{Spherically averaged structure function for parameter set
III ({\em cf.} Table \ref{PARAMETERS}). $128^3$ lattice. Error bars
represent the standard error of the structure function spherical
average. Time slices shown are time step 100, 200, 300, 400, 500, 600,
and 700 from right to left. All quantities are reported in lattice
units.}   
\label{VISC128_SF_vs_k}
\end{center}
\end{figure}

Analogous behaviour to Fig. \ref{DIFF128_SF_vs_k} is exhibited for
parameter sets II, III, and IV ({\em cf.} Table \ref{PARAMETERS}) in
Figs. \ref{DIFFB128_SF_vs_k}, \ref{VISC128_SF_vs_k}, and
\ref{INER128_SF_vs_k}, respectively. For the last two time slices
taken in Fig. \ref{INER128_SF_vs_k}, the peaks seem no longer to drift
to the left, as a result of finite size effects (arrest of domain
growth). Regarding regions of exponential growth with time, the three
data sets confirm Eq. (\ref{EXP_GROWTH}), with an upper bound for $k$.

%% COLLAPSED SF PLOTS

\begin{figure}[p]
\begin{center}
\includegraphics[angle=0,width=16cm]{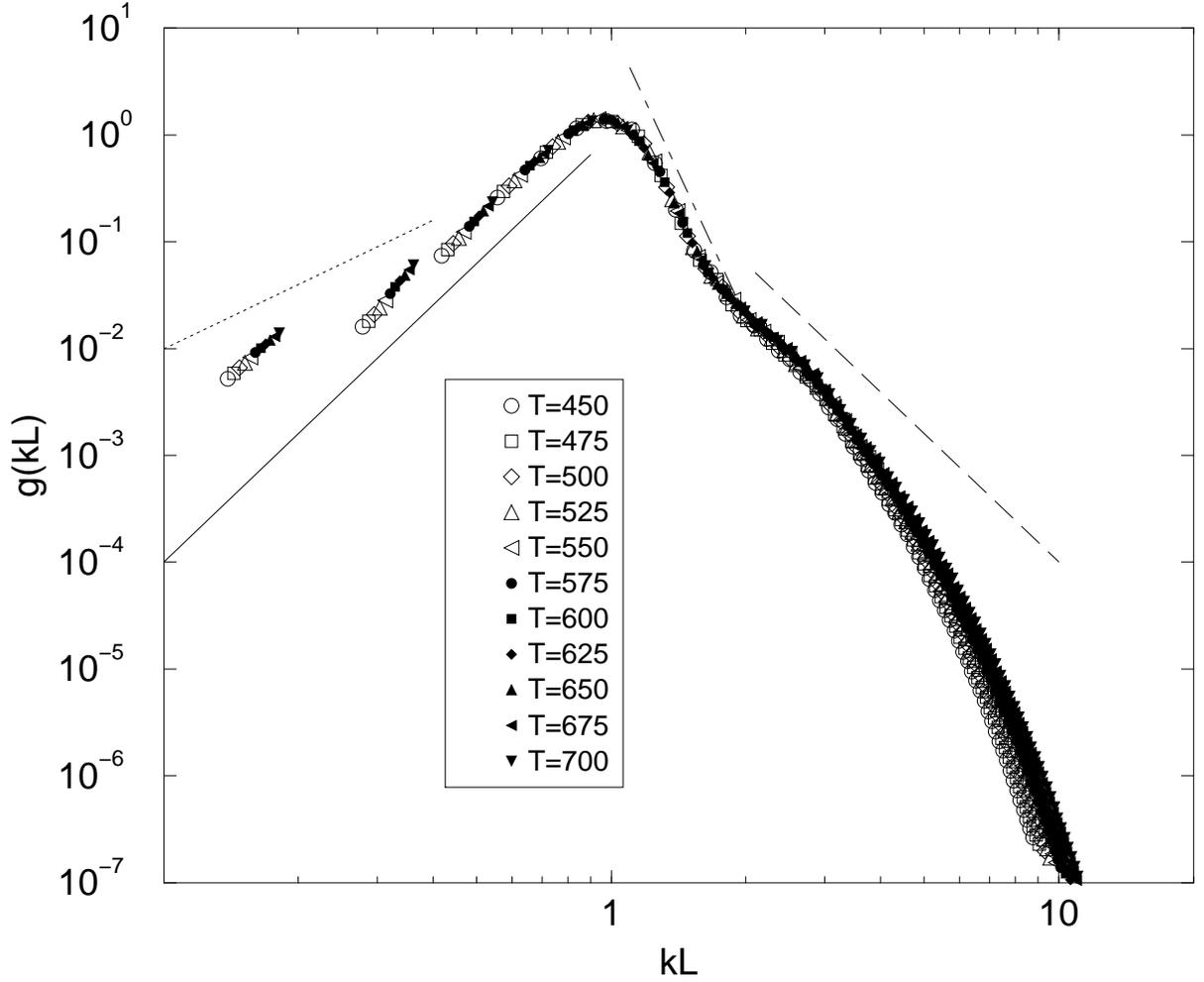}
\caption{Scaled spherically averaged structure function for parameter set
III ({\em cf.} Table \ref{PARAMETERS}), as defined by
Eq. (\ref{COLLAPSED-SF}). Lattice size is $128^3$. Time steps are as
shown in the legend. Earlier times correspond to the empty symbols;
later times to the filled symbols. Error bars are smaller than the
size of the symbols. Straight lines serve as slope guides to the
reader only, and represent power laws $q^2$, $q^4$, $q^{-9}$, and
Porod's law $q^{-4}$, from left to right, respectively, with 
$q\equiv kL$. All quantities are reported in lattice units.}   
\label{VISC128_SCSF}
\end{center}
\end{figure}

\begin{figure}[p]
\begin{center}
\includegraphics[angle=-90,width=16cm]{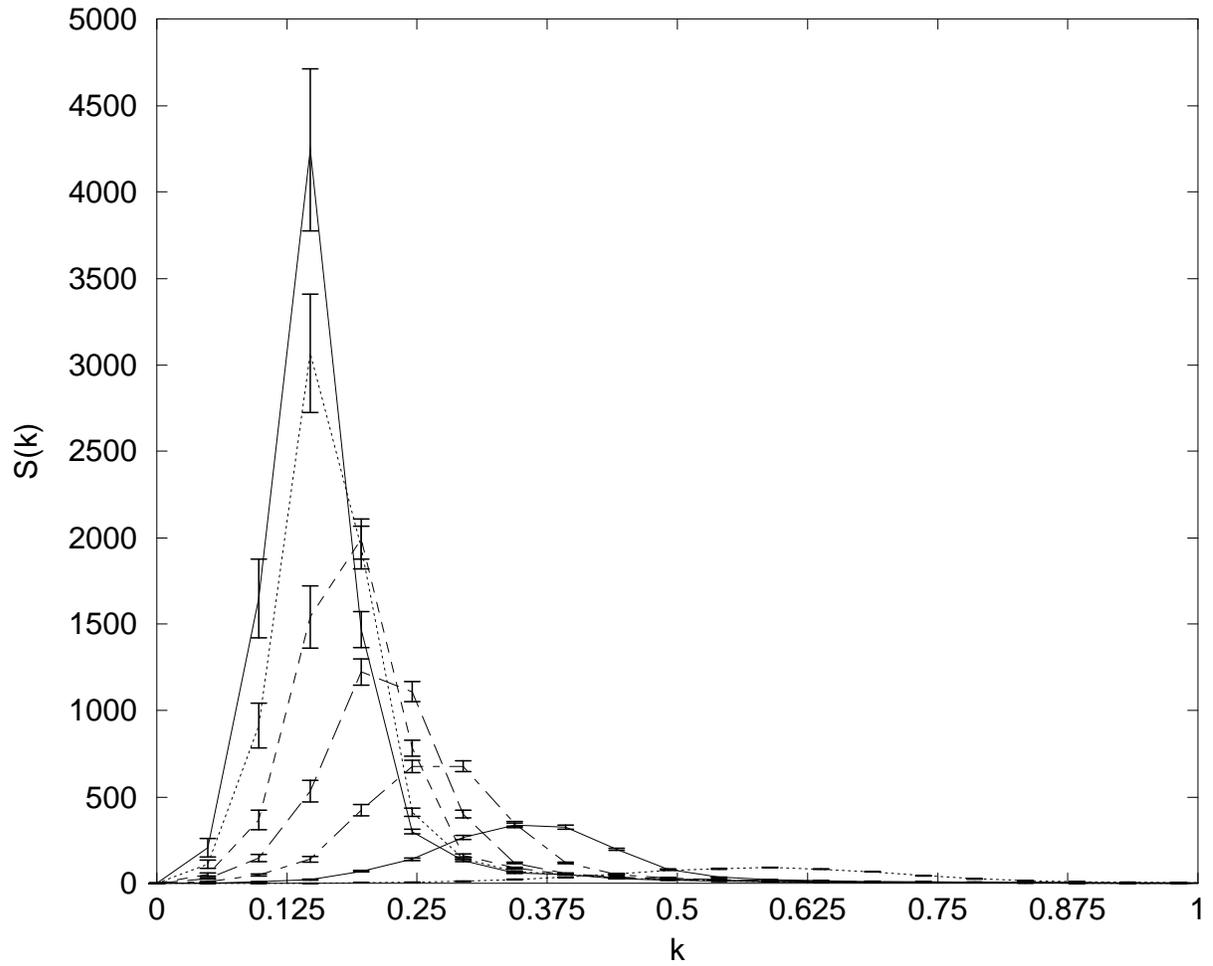}
\caption{Spherically averaged structure function versus wavenumber,
for parameter set IV ({\em cf.} Table \ref{PARAMETERS}). $128^3$
lattice. Error bars represent the standard error of the structure
function spherical average. Time steps shown are 100, 200, 300, 400,
500, 600, and 675, from right to left. All quantities are reported in
lattice units.}  
\label{INER128_SF_vs_k}
\end{center}
\end{figure}

Figure \ref{VISC128_SCSF} shows the collapse (matching) of the
structure functions corresponding to parameter set III ({\em cf.}
Table  \ref{PARAMETERS}), for a $128^3$ lattice size and time steps
from 450 to 700, when they are scaled by Eq. (\ref{COLLAPSED-SF}), the
abscissae being rescaled by a factor of $(2\pi)^{-1}$, and the
ordinates by the peak's maximum. Earlier times are represented in
Fig. \ref{VISC128_SCSF} by empty symbols, and later times by filled
symbols. There is good collapse, and, therefore, scaling according to
the scaling hypothesis, in the region from $q=0.4$ to about
$q\approx 3$, where $q\equiv kL$ is dimensionless. The middle of the
region $1<q<2$ follows a $q^{-9}$ behaviour, in accordance with
Tomita's prediction of an exponent $-6$ or more negative \cite{TOMITA}. 

Close to $q\approx 3$ we observe the presence of a shoulder, as has
been reported in experiments \cite{KUBOTA} and numerical simulations 
\cite{APPERT-LGASD,KOGA}. Most strikingly, the shape of 
our large-$q$ tail is very reminiscent of that of Fig. 4 in
\cite{APPERT-LGASD} and that of Fig. 3 in \cite{KOGA}: (1) there is
still a time dependence indicating that interfaces have not yet been
fully resolved (we are probing the smallest scales where diffusion
still exists and $\xi/L$ is not small enough); and (2) the tail
decreases with an exponent which is in fact more negative than that of
the Porod tail, Eq. (\ref{POROD}), despite what these authors
\cite{APPERT-LGASD,KOGA} claim. 

For $q<0.4$, data points do not seem to collapse onto the same curve
of those for $q>0.4$. This is similar to, but with more data than, the 
results of Koga and Kawasaki \cite{KOGA}.  Our results show an
exponent growing with time: the slope of a line (not shown) joining
the first two empty circles ($T=450$) is 1.61, while the slope of a
line (not shown) joining the last two filled downward triangles
($T=700$) is 2.12. This resembles the temporal growth cited by Appert 
{\it et al.} \cite{APPERT-LGASD} on the results of Alexander {\it et
al.} \cite{ALEXANDER93}; nonetheless, we consider the amount of data
in the latter insufficient to draw firm conclusions. Given that the
points at $T=700$ are closer to the asymptotic regime, we take such a
slope as our best approximation to the asymptotic regime.

In the small-$q$ region, Yeung \cite{YEUNG} predicted a $q^4$ behaviour
for the asymptotic limit ($L\to\infty$, or at late
times). Additionally, at earlier stages, a term proportional to   
$L^{-2}q^2$ caused by thermal noise would also come into play. Now,
Appert {\it et al.}'s estimate \cite{APPERT-LGASD} applies well for
our results: such a quadratic term is less dominant than the quartic
one only for $q>0.4$, given that the largest value of $L(T)$ for which
there are no finite size effects is also about 25. This happens to be
the region where we find the $q^2\leftrightarrow q^4$ crossover. 

\begin{figure}[p]
\begin{center}
\includegraphics[angle=-90,width=16cm]{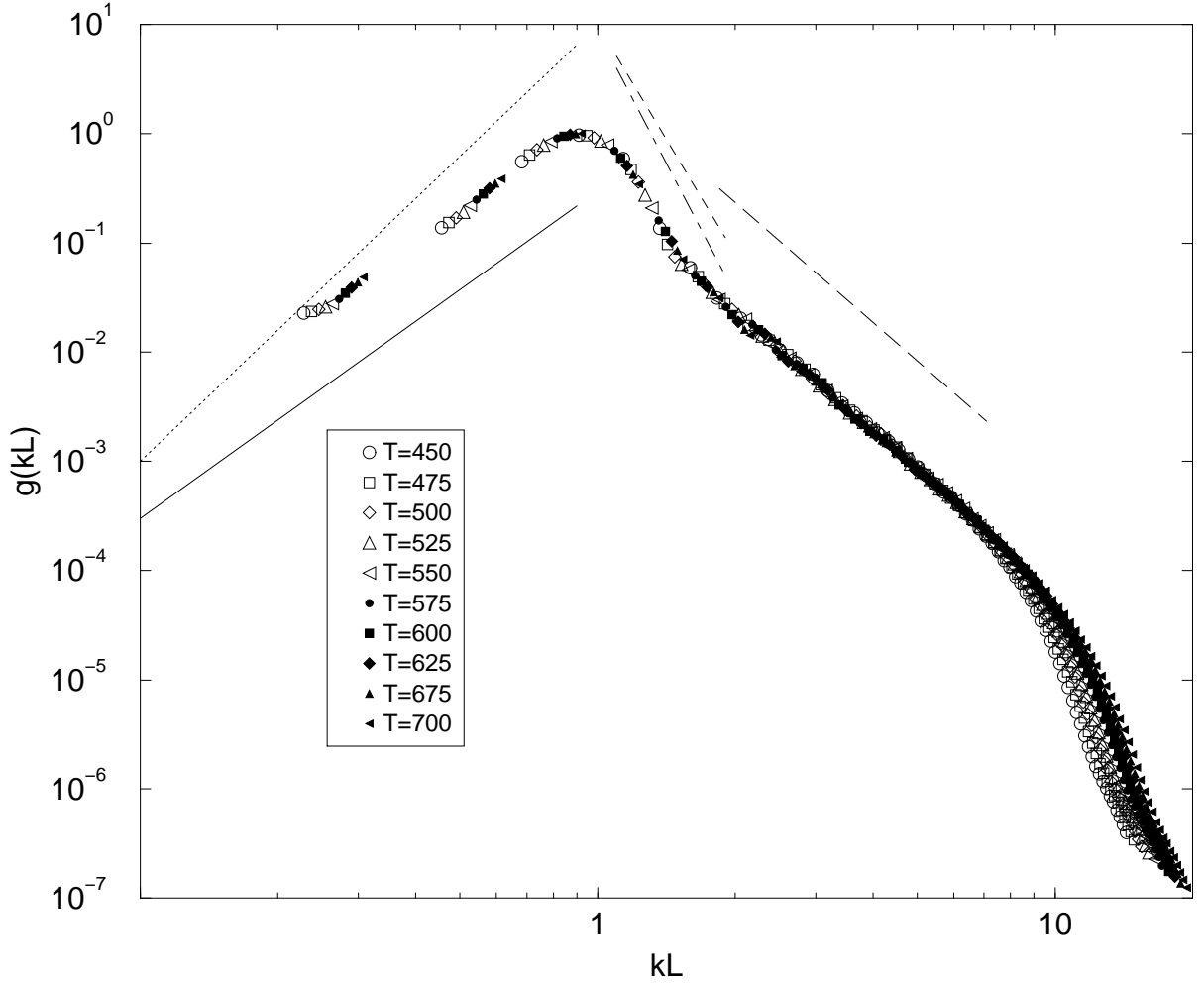}
\caption{Scaled spherically averaged structure function for parameter
set IV ({\em cf.} Table \ref{PARAMETERS}), as defined by
Eq. (\ref{COLLAPSED-SF}). Lattice size is $128^3$. Time steps shown
are from 450 up to 675, every 25. Earlier times correspond to the
innermost lines; later times to the outermost lines. Error bars are
smaller than the size of the symbols, except for the two leftmost,
detached data sets, for which they are slightly larger. Straight
lines serve as slope guides to the reader only, and represent power
laws $q^3$, $q^4$, $q^{-8}$, $q^{-7}$, and the fit to the large-$q$
tail, $q^{-3.65}$, from left to right, respectively, with $q\equiv
kL$. All quantities are reported in lattice units.} 
\label{INER128_SCSF}
\end{center}
\end{figure}

Fig. \ref{INER128_SCSF} shows similar curves for parameter set IV
({\em cf.} Table \ref{PARAMETERS}), where only time steps 450 to 675
are displayed and we have also normalised the curve such that the peak
is located at (1,1). We have again neglected early time steps because
of poor collapse. A fit to the tail in $2<q<10$ gives $q^{-3.65}$,
close to being a Porod's law. It is when we probe the finest length
scales, at $q\approx 10$ that it ceases to apply, due to lattice
discretisation effects. 

The behaviour at intermediate wavenumbers is between $q^{-8}$ and
$q^{-7}$, again in agreement with Tomita's theory \cite{TOMITA}, and
close to $q^{-7}$ as computed using a dissipative particle dynamics
method by Jury {\it et al.} \cite{JURY99} and a lattice-gas automaton 
by Love {\it et al.} \cite{LOVE}.

For small momenta (large domains) we found a behaviour close to $q^3$,
in agreement with the numerical results of Love {\it et al.}
\cite{LOVE} and in disagreement with Yeung's predictions
\cite{YEUNG}. 

The most notable difference between Figs. \ref{VISC128_SCSF} and
\ref{INER128_SCSF} is the behaviour above $q\approx 1.5$. Figure
\ref{INER128_SCSF} shows a neat Porod tail, which bends down
dramatically for $q>10$, whereas Fig. \ref{VISC128_SCSF} shows either
a poor Porod tail in the region $3<q<5$, or a minute one in the
region $1.5<q<3$. A condition assumed in the derivation of Porod's law
\cite{BRAY} is that the sampling length $r$ satisfies $\xi \ll r \ll
L$, which in wavenumbers means

\begin{equation}
	1/L \ll k\ll 1/\xi	\,.  \label{k_RANGE_SCALING}
\end{equation}

\begin{figure}[p]
%\begin{center}
\includegraphics[angle=0,width=10cm]{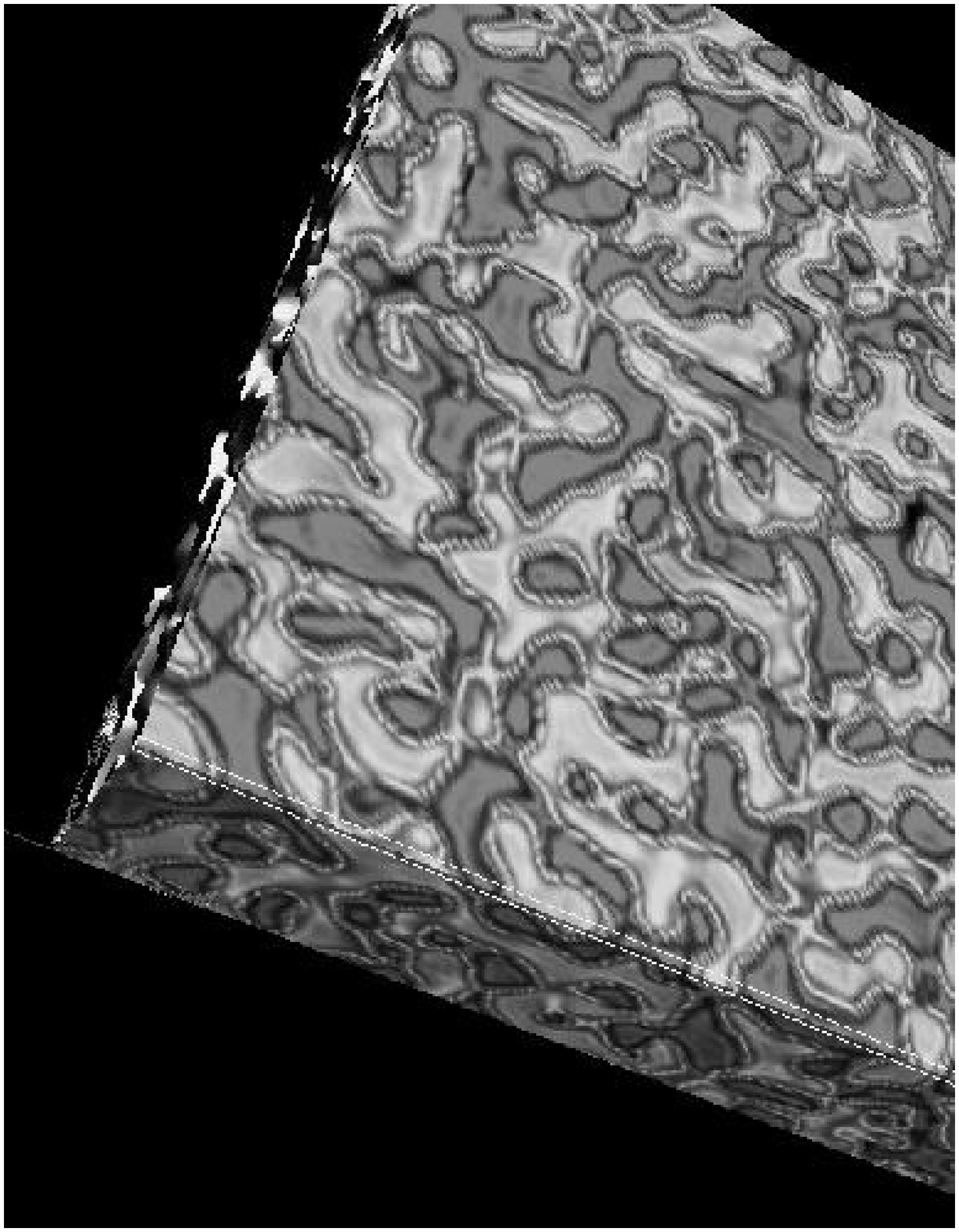}
\caption{Order parameter ($\rho^R-\rho^B$) snapshot at time step 200
for parameter set IV ({\em cf.} Table \ref{PARAMETERS}). We show a
slab of the $256^3$ lattice simulated.}   
\label{PIC_INER256_t000200}
% \end{center}
\end{figure}

\begin{figure}[p]
\begin{center}
\includegraphics[angle=0,width=16cm]{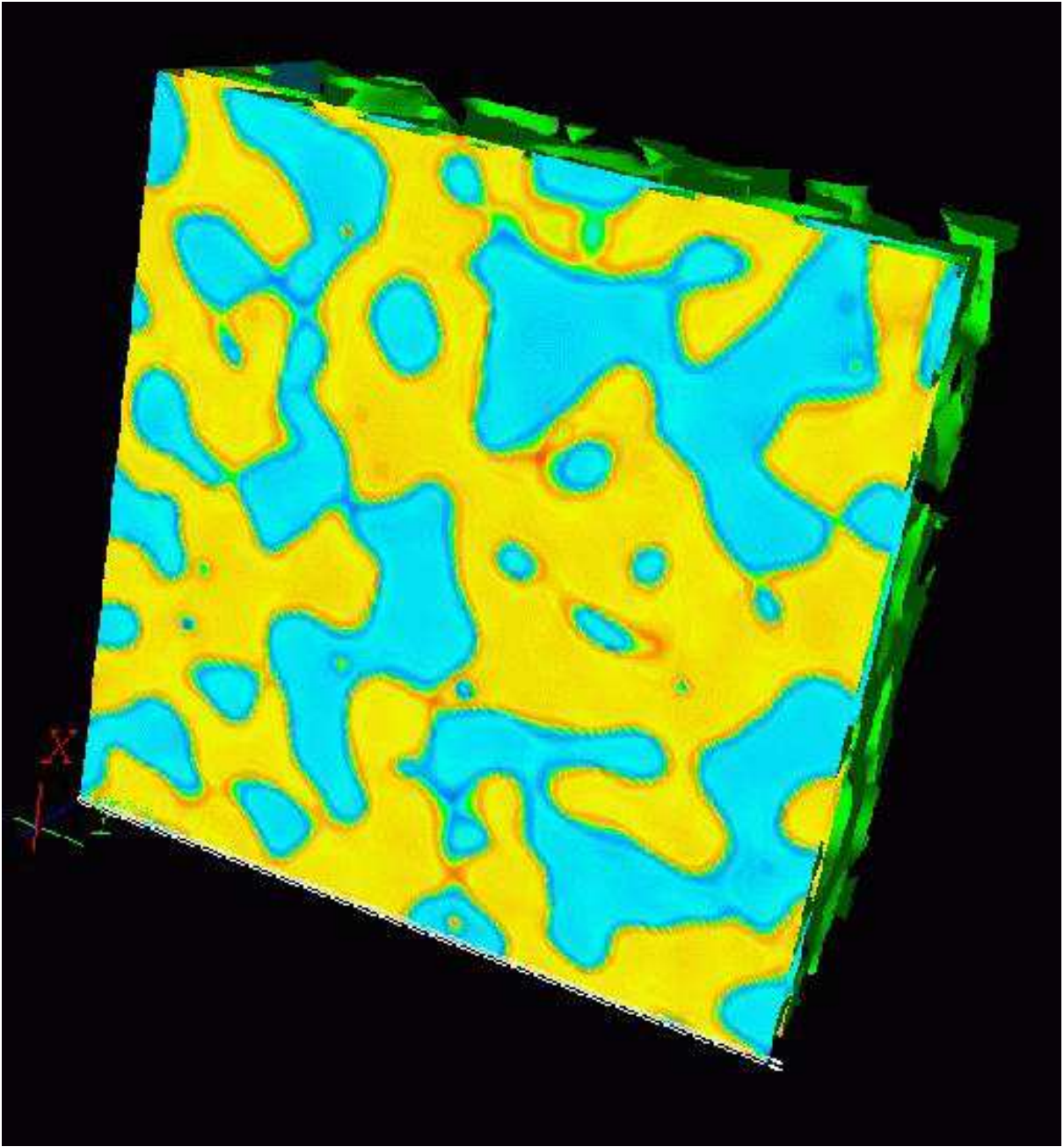}
\caption{Order parameter ($\rho^R-\rho^B$) snapshot at time step 700
for parameter set IV ({\em cf.} Table \ref{PARAMETERS}). We show a
slab of the $256^3$ lattice simulated.}   
\label{PIC_INER256_t000700}
\end{center}
\end{figure}

\noindent By `eyeball' inspection of the system's order parameter we
found that interface widths naturally shrink with an increasing
number of time steps, going from about 5 or 6 lattice unit spacings at
200 time steps down to about 3 at 675 time steps, regardless of the
data set, III or IV ({\em cf.} Table \ref{PARAMETERS}). Simulations
for a $256^3$ lattice size revealed similar widths, and snapshots
of the order parameter at 200 and 700 time steps are shown in
Figs. \ref{PIC_INER256_t000200} and \ref{PIC_INER256_t000700}. With
these widths in mind, assuming domain sizes of a quarter of the
lattice side length (the threshold imposed by our prescription for
eliminating finite size effects), and a $128^3$ lattice, the condition
(\ref{k_RANGE_SCALING}) becomes

\begin{equation}
	1 \ll q \ll 10	\,, \label{q_RANGE_SCALING}
\end{equation}

\noindent which contains our large-$q$ region. Despite this, we do not
observe a Porod tail for data set III, or, as in data set IV, the tail
obtained is only close to being a Porod tail. This is in agreement
with the fact Eq. \ref{k_RANGE_SCALING} is necessary but not
sufficient for a Porod tail to hold.

Finally, it is worth noting in
Fig. \ref{PIC_INER256_t000700} the existence of nested domains and
droplets much smaller than the average domain size.

%%%%%%%%%%%%%%%%%%%%%%%%%%%%%%%%%%%%%%%%%%%%%%%%%%%%%%%%%%
%% P L O T S     A N D     P I C T U R E S
%%%%%%%%%%%%%%%%%%%%%%%%%%%%%%%%%%%%%%%%%%%%%%%%%%%%%%%%%%

%%% STRUCT FUNCT: 

%%%%%%% DIFF128

%%%%%%% DIFF128: Family of S-vs-T curves, for different k
%%%%%%%  

%%% STRUCT FUNCT: 

%%%%%%% DIFFB128

%%% STRUCT FUNCT: 

%%%%%%% VISC128

%%% STRUCT FUNCT: 

%%%%%%% INER128

%%% PICTURES: INER256

\section{Numerical stability of our lattice-Boltzmann algorithm}

As is well known, owing to the lack of an H-theorem, an approach to
equilibrium is not guaranteed in all lattice-Boltzmann models to date;
recent theoretical developments to address and solve this have been
made \cite{KARLIN,CHEN-TEIXEIRA,BOGHOSIAN-LBE}. For single-phase
lattice-Boltzmann models, equilibrium states are well defined in the
collision term; if the automaton does relax to these, the pertinent
macroscopic momentum (and sometimes energy) balance equations are
reproduced in the low-Knudsen number limit. Interacting,
multicomponent lattice-Boltzmann models exhibit the same situation in
the bulk of pure fluid regions where intercomponent interactions are 
negligible. For regions where they are not, there is not even a
well-established thermohydrodynamic theory which could provide
equilibria to which the automaton could relax to, or with which to
compare the stationary state to which it can evolve. Whether
dealing with a single or multiphase system, lattice-BGK stationary
regimes ought to be treated with caution and contrasted with
experiment.

Numerical instabilities are the reflection of the lack of an H-theorem,
which is a direct consequence of space and time discretisation on the
BGK-Boltzmann equation and the freedom in the choice of the
equilibrium distribution function \cite{BOGHOSIAN-LBE}. These 
instabilities can be defined as follows. As is generally the case for
a finite difference method with a single relaxation parameter, such as
our lattice-BGK model for a zero phase-coupling constant \cite{QIAN},
linear stability occurs within a finite interval of such a
parameter. If multicomponent interactions are introduced, additional
parameters may influence the stability: density, intercomponent
coupling strength, and even composition. The mechanism is simple:
certain choices of parameters can turn the lattice-BGK collision term
positive (therefore increasing the mass density) for long enough to
generate floating-point numbers larger than the largest the machine
can deal with, hence causing an overflow signal. Numerical
instabilities are defined in this work as the generation of such
floating-point numbers. We consider it crucial to be able to map out
regions in the model's parameter space leading to unstable
configurations, and to report them alongside any lattice-BGK
simulations.   
                                                      
Using the same initial conditions as explained in Section \ref{SIMUL},
we found our algorithm to be unstable for regimes with 
the smallest length and time scales, $L_0$ and $T_0$,
which coincide with those of the largest Mach
numbers. In Table \ref{INSTABILITY_PARMS} we show some of the
paramaters leading to numerical instability. The dependence of the
surface tension on the model parameters, as given in Section 
\ref{SURFTENS}, should be taken into consideration as a guide to 
steering through the parameter space. Note that all values of $\Delta t$
included are larger than that for parameter set IV.

\begin{table}[!htb]
\begin{center}
\begin{tabular}{|c|c|c|c|c|}
\hline
$\rho$ & $\tau$ & $g$ & $\sigma(\rho,\tau,g)$ & $T_0$ \\ 
\hline
0.5    & 0.5625   & 0.06 & 0.0115  & 0.0169  \\   
0.5    & 0.5625   & 0.03 & 0.0052  & 0.0122  \\
0.3    & 0.5625   & 0.10 & 0.0068  & 0.0174  \\   
0.3    & 0.5500   & 0.08 & 0.0061  & 0.0112  \\
\hline
\end{tabular}
\end{center}
\caption{\small Model parameters leading to numerical instability,
including the surface tension $\sigma(\rho,\tau,g)$ generated some
time steps before the instability sets in, and the associated
characteristic time. The lattice used was $4\times4\times128$, and the
instability sets in before 4000ts.}
\label{INSTABILITY_PARMS}
\end{table}

\begin{figure}[p]
\begin{center}
\includegraphics[angle=-90,width=16cm]{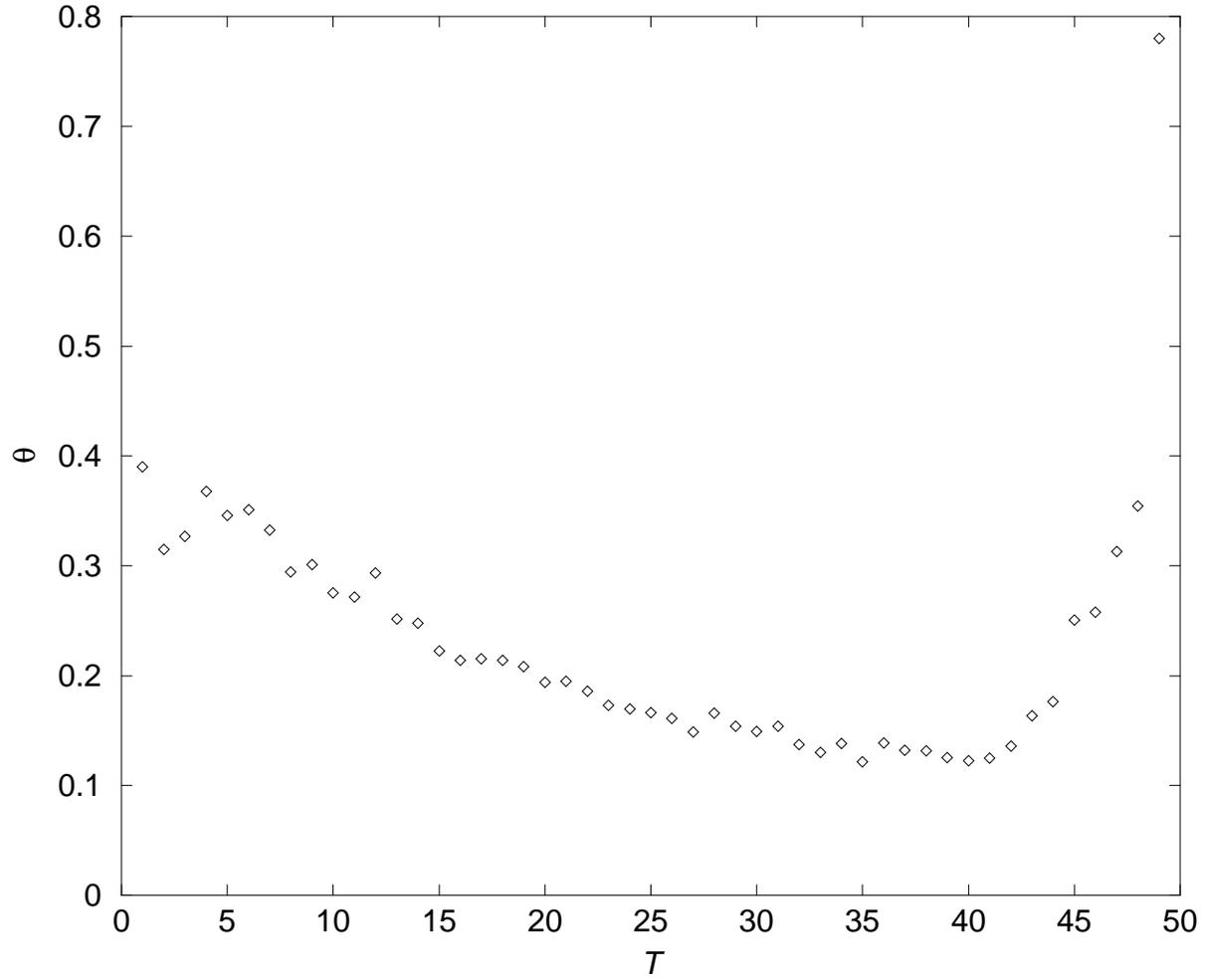}
\caption{Evolution of the collision term maximum absolute value,
$\theta$, Eq.(\ref{THETA}), with the time step on a $32^3$ lattice for
parameters $\{\rho=0.3,\,g=0.06,\,\tau=0.5125\}$. The interpolating
curve serves as a guide to the eye only. All quantities are in lattice
units.}  
\label{THETA_UNSTAB32}
\end{center}
\end{figure}

\begin{figure}[p]
\begin{center}
\includegraphics[angle=0,width=16cm]{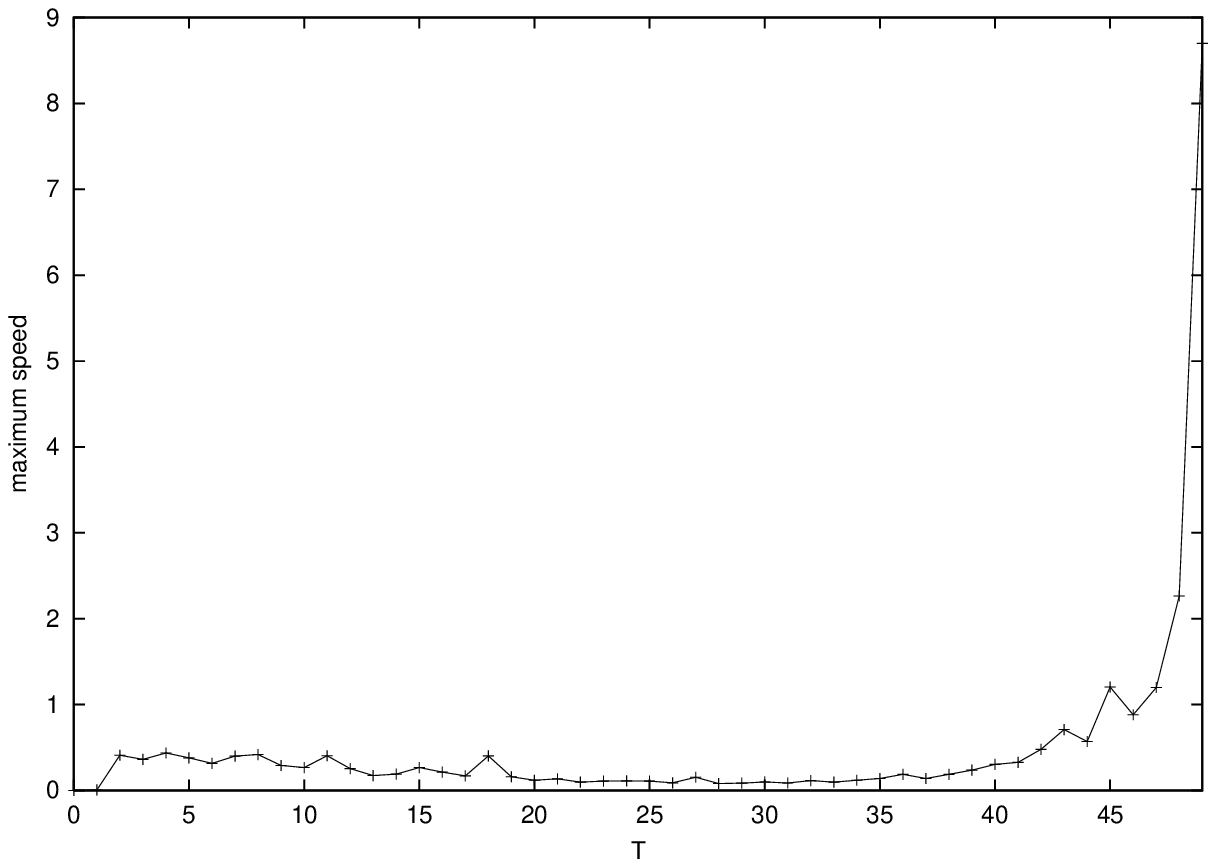}
\caption{Evolution of the maximum speed, $u_{\mathrm{max}}$, with the
time step on a $32^3$ lattice for parameters
$\{\rho=0.3,\,g=0.06,\,\tau=0.5125\}$. The interpolating curve serves 
as a guide to the eye only. All quantities are in lattice units.}    
\label{VELMAX_UNSTAB32}
\end{center}
\end{figure}

\begin{figure}[p]
\begin{center}
\includegraphics[angle=0,width=16cm]{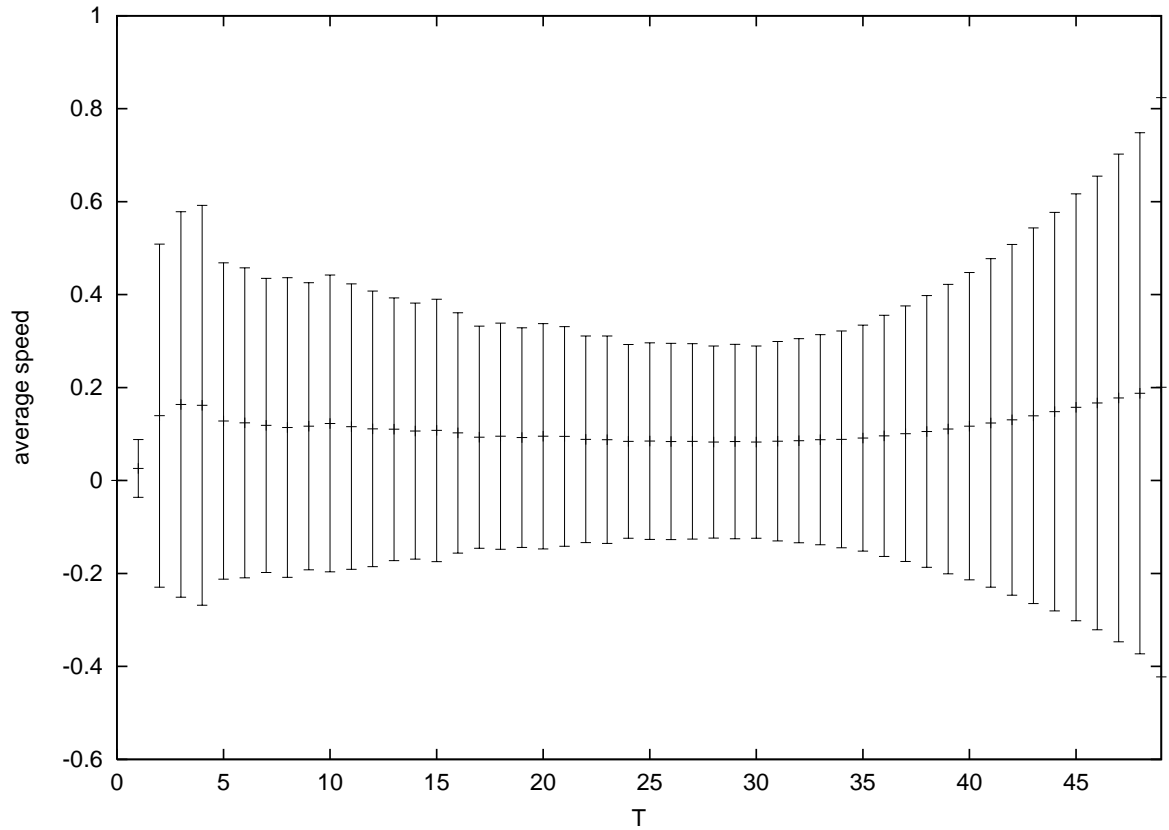}
\caption{Evolution of the speed average, $\overline u$, with the time
step on a $32^3$ lattice for parameters
$\{\rho=0.3,\,g=0.06,\,\tau=0.5125\}$. Error bars represent the  
standard error of the average (``one sigma''). All quantities are in
lattice units.}
\label{VELAVG_UNSTAB32}
\end{center}
\end{figure}

\begin{figure}[p]
\begin{center}
\includegraphics[angle=-90,width=16cm]{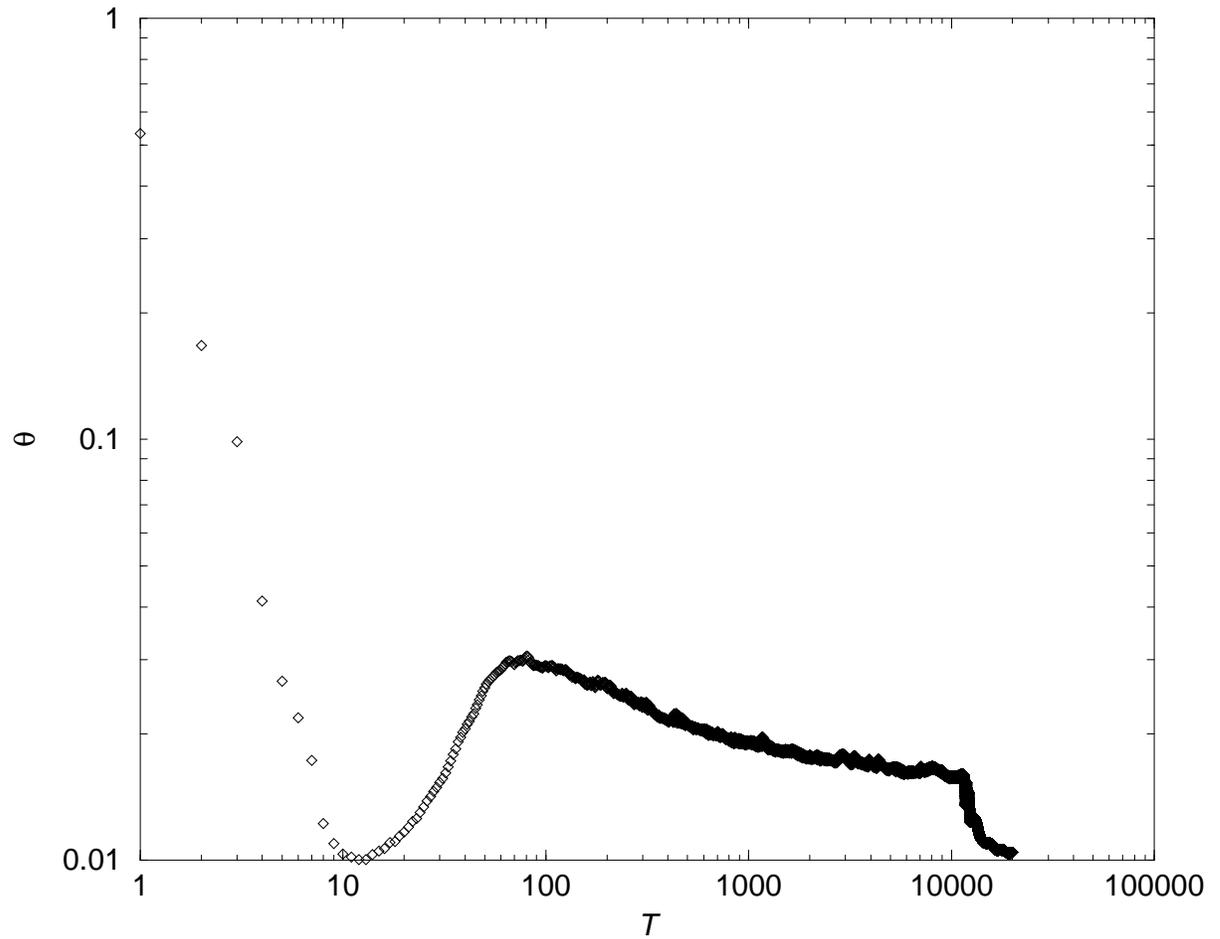}
\caption{Evolution of the collision term maximum absolute value,
$\theta$, Eq.(\ref{THETA}), with the time step on a $128^3$ lattice
for parameter set I (cf. Table \ref{PARAMETERS}). We can see a
decreasing trend for most of the simulation, which accentuates after
time step 10000. The interpolating curve serves as a guide to the eye
only. All quantities are in lattice units.}   
\label{THETA_STAB128}
\end{center}
\end{figure}

\begin{figure}[p]
\begin{center}
\includegraphics[angle=0,width=16cm]{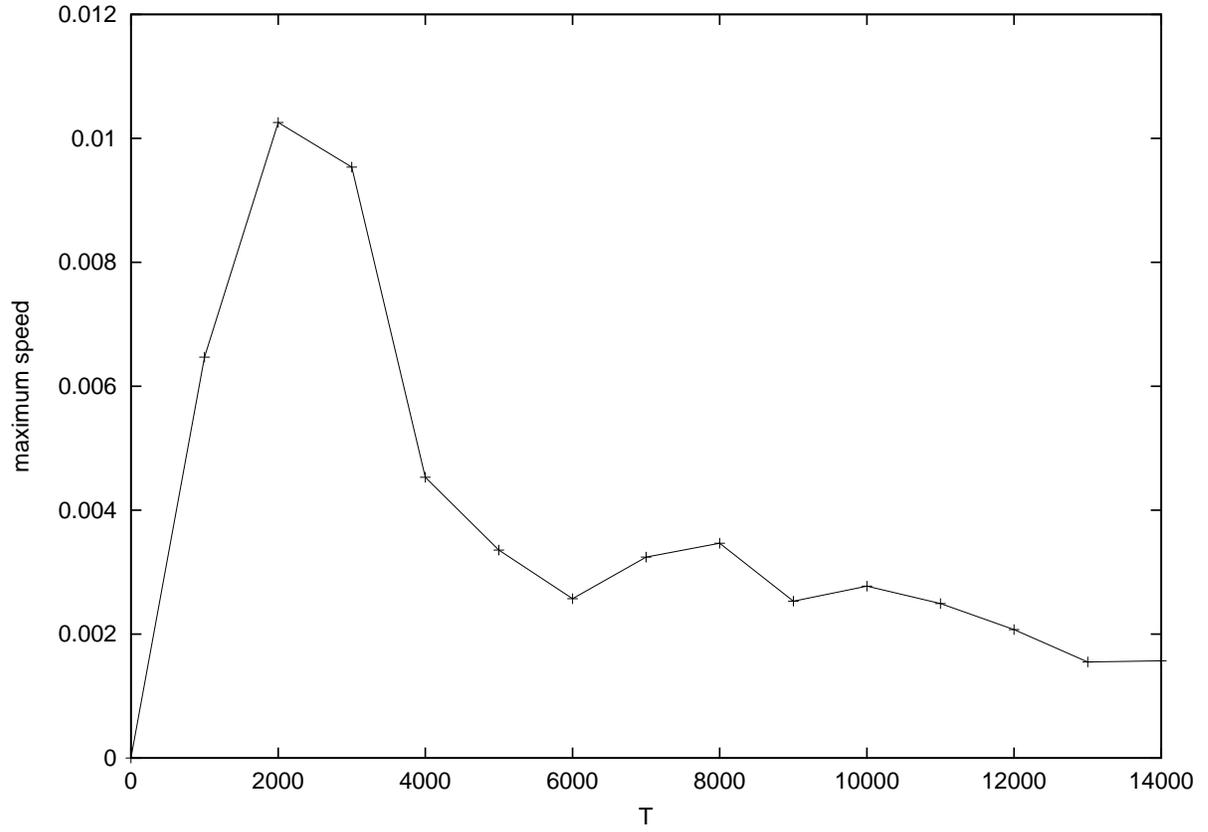}
\caption{Behaviour of the maximum speed, $u_{\mathrm{max}}$, with the
time step for a $128^3$ lattice with parameter set I  
(cf. Table \ref{PARAMETERS}). It shows an overall decreasing
trend. The interpolating curve serves as a guide to the eye only. All
quantities are in lattice units.}  
\label{VELMAX_STAB128}
\end{center}
\end{figure}

\begin{figure}[p]
\begin{center}
\includegraphics[angle=0,width=16cm]{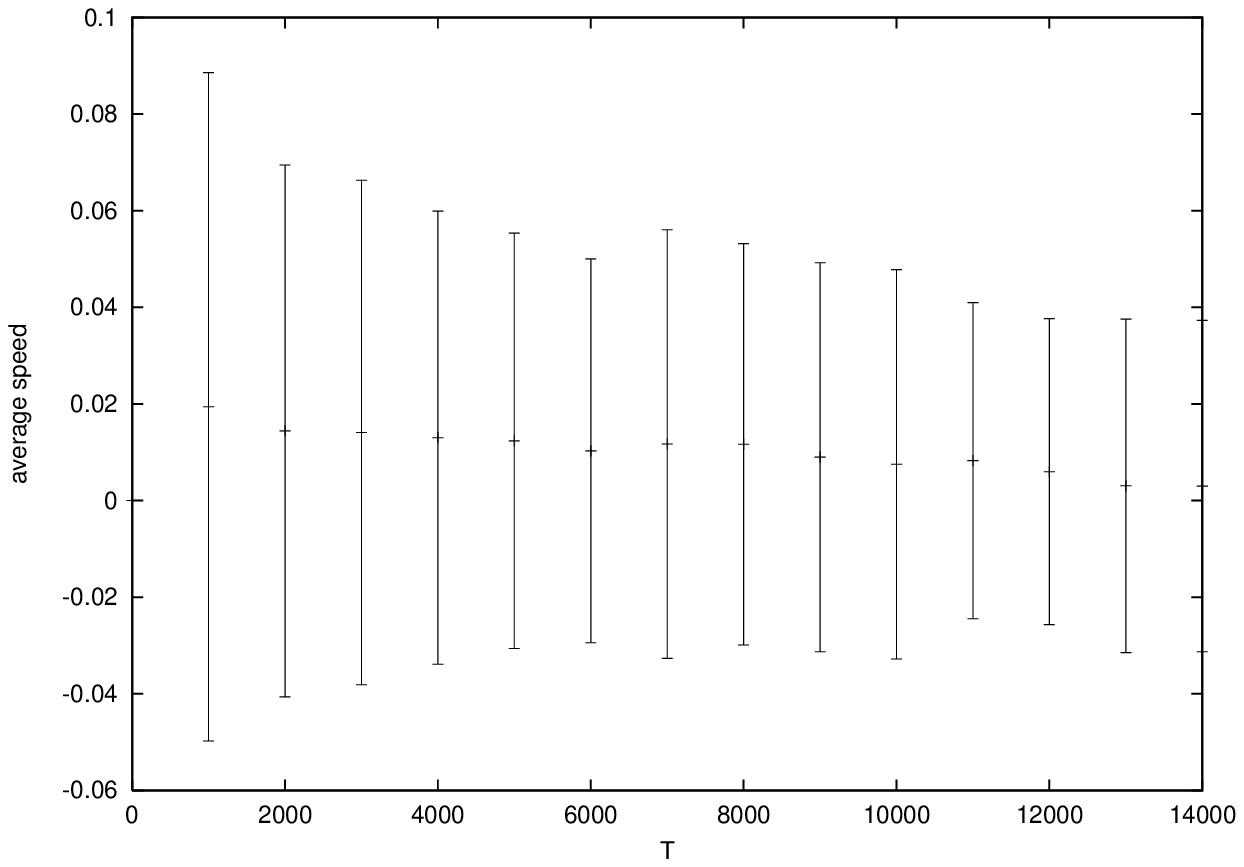}
\caption{Evolution of the speed average, $\overline u$, with the time
step on a $128^3$ lattice for parameter set I (cf. Table
\ref{PARAMETERS}). We can see a decreasing trend of the average and 
its error. All quantities are in lattice units.}  
\label{VELAVG_STAB128}
\end{center} 
\end{figure}

We then investigated the nature of our instabilities, as others have
done. The group of Cates found troublesome numerical
instabilities with their free-energy based, lattice-BGK model in 3D in
regions in which quiescent binary portions of fluid go into a 
checkerboard state \cite{CATES_PRIVATE}. They reported that
their model is unconditionally unstable \cite{KENDON01}. Nonetheless,
by improving the way gradients were treated numerically they were able
to considerably reduce this unphysical behaviour. For our model, we
looked at the time evolution of the quantity 

\begin{equation}
	\theta(t)\equiv
	\max\{|\Omega_k^{\prime\alpha}({\mathbf x},t)|\,,\,\,
	\forall {\mathbf x},\forall k,\forall\alpha\}\,,
	\label{THETA}
\end{equation}

\noindent for parameters $\{\rho=0.3,\,g=0.06,\,\tau=0.5125\}$, where
the collision term, $\Omega_k^{\prime\alpha}$, is defined in
Eq. (\ref{LBEQUATION}). We also monitored the maximum and average
values of the fluid mixture's speed, $u_{\mathrm{max}}$ and
$\overline u$, respectively, on the lattice. We show these 
quantities for a $32^3$ lattice in Figs. \ref{THETA_UNSTAB32},
\ref{VELMAX_UNSTAB32} and \ref{VELAVG_UNSTAB32}. We see how $\theta$
reverses its decreasing trend in a few time steps; after that, it
blows up at $T=52$ time steps. We only show data up to $T=49$, as
$\theta(T=51)\approx 10^{111}$. $u_{\mathrm{max}}$ blows up in similar
style: at $T=50$, $u_{\mathrm{max}}=7498$ and $\theta\approx 10^{21}$;
at $T=51$, $u_{\mathrm{max}}$ has exceeded the maximum floating-point
value that the computer can deal with, and overflow signals are
generated. This indicates that at the time steps immediately prior to
the onset of the instability the lattice gets more and more populated
with increasing speeds until in two or three time steps they
grow by ten or more orders of magnitude. That the population of
lattice sites with rapidly increasing speeds over time is
small compared to the lattice volume can be concluded from contrasting
the time variation in the standard error (``one sigma'') of
${\overline u}$ to the time variation of ${\overline u}$,
Fig. \ref{VELAVG_UNSTAB32}. The same parameter set run on a $128^3$
lattice seems to make the instability set in much quicker, as it
occurs during the first 10 time steps. As a final check, we ran a
$128^3$ lattice with parameter set I (cf. Table \ref{PARAMETERS}) for
20000 time steps and found no instabilities. The time evolution of
$\theta$, $u$ and $\overline u$ is shown in Figs. \ref{THETA_STAB128},
\ref{VELMAX_STAB128} and \ref{VELAVG_STAB128}, respectively. We
conclude that the occurrence of instabilities only depends on the set
of parameters used, regardless of the number of time steps simulated.

%%%
%%% STABILITY

%%% UNSTABLE 32^3

%%% STABLE 128^3

\section{Conclusions}
\label{CONCLU}

We have presented a quantitative study of the phase separation
dynamics in three dimensions for critical (50:50) fluid mixtures
(spinodal decomposition) for a modified Shan-Chen lattice-BGK model
of multicomponent, isothermal immiscible fluids. 

We found that after a brief diffusional transient in which
interconnected regions of fluid species embedded into one another are
formed, the average size of such regions grows with time as $l\propto 
t^\gamma$, where $\gamma\approx 2/3$. The trend is for the value to
increase in the range $0.545\pm 0.014<\gamma<0.717\pm 0.002$ as the
Reynolds number increases. This increase is consistent with a
crossover from $l\propto t^{1/3}$ diffusive behaviour to hydrodynamic
viscous growth $l\propto t$ predicted by the Cahn-Hilliard Model
H. Owing to the significant amount of diffusion present at low
Reynolds number, we do not consider our results to be indicative of a
genuinely hydrodynamic inertial $t^{2/3}$ regime.

We observed exponential growth in the time dependence of the
structure function for wavenumbers up to a threshold value, in
qualitative agreement with predictions from the linearised
Cahn-Hilliard Model B. For small wavenumbers, such an exponential
growth is seen at all simulation times, whereas it is only an
initial transient for larger wavenumbers. These departures from
Model B predictions are for wavenumbers far from the one
characterising the average domain size. A natural continuation of
this work would be to investigate the nature of diffusion currents
for these cases. 

We have found very good agreement with the dynamical scaling
hypothesis in the form of a neat collapse of the structure
function curves for $\mathrm{Re}=2.7$ and $\mathrm{Re}=37$ when they
are appropriately scaled according to Eq. (\ref{COLLAPSED-SF}). This
collapse holds roughly for the second half of the simulation time, as  
diffusional transients act during the first. By looking at order
parameter snapshots we observed the formation of nested domains and
smaller droplets for the largest Reynolds numbers achieved, as Wagner
and Yeomans also found \cite{WAGNER98}. However, unlike them, in our
case these are transients rather than a result of length scales
growing at different speeds, as poor collapse of the scaling functions
would then occur due to breakdown of scale invariance.  

Yeung predicted a $q^2$ behaviour at the small-$q$ end of the spectrum
as the result of thermal effects at pre-asymptotic stages
\cite{YEUNG}. Love {\it et al.} \cite{LOVE} conjectured that a $q^2$
behaviour, and a crossover to $q^4$, could be caused by (a)
lattice-gas noise, or (b) a poor scaling collapse, and that their 
$t^{2/3}$ domain growth, instead of $t$, might be justified by the
former. Appert {\it et al.} \cite{APPERT-LGASD} ascribed the $q^2$
behaviour and the crossover to not having reached the asymptotic
limit, $L\to\infty$ (poor scaling collapse again). Our noiseless model
reproduced the $q^2\leftrightarrow q^4$ crossover at $\mathrm{Re}=2.7$
and did not at $\mathrm{Re}=37$, for which there is better scaling
collapse, and also produced a 2/3 domain growth (crossover)
exponent. All this leads us to conclude that noise may not play as
important a role as the lack of scaling collapse in explaining 
the $q^2\leftrightarrow q^4$ crossover, and is definitely not a 
requirement for the reproduction of a 2/3 domain growth exponent.
A $q^2$ behaviour is the only one experimentally observed by Kubota
{\it et al.} \cite{KUBOTA} for a mixture of isobutyric acid and water;
they cite surface tension effects, measurement difficulties, multiple
light scattering, and even specificity to the mixture's molecular
weight as reasons for not seeing a $q^4$ behaviour, and definitely
discard thermal noise. Not surprisingly, in his prediction Yeung
assumed a diffusive domain growth exponent of $1/3$, which is rather
seen in quenches of polymer mixtures and metal alloys. 

In the case $\mathrm{Re}=37$, the spectrum shows a $q^3$ behaviour in
the small-$q$ limit, in disagreement with Yeung's prediction. In fact,
his analysis is based on a Cahn-Hilliard model without hydrodynamics.

The numerical instabilities seen in our runs are caused by large
speeds turning the equilibrium distribution negative for long enough
to incur floating-point overflows. This happens for characteristic
times ({\em cf.} Table \ref{PARAMETERS}) below $T_0=0.0172$, and the
population of lattice sites undergoing such a burst in the fluid's
macroscopic speed is small compared to the lattice volume. We found no
evidence that an initially stable regime becomes unstable at later
times, as typically happens in relaxational models (such as is our
model for $g_{\alpha\overline\alpha}=0$). This is in stark contrast
with the findings of Kendon {\em et al.} \cite{KENDON01} and Cates
{\em et al.} \cite{CATES_PRIVATE} in their spinodal decomposition
studies using a free-energy based, lattice-BGK model, who reported
their algorithm to be unconditionally unstable.  

A search for a crossover to growth laws other than $t^{2/3}$ at
Reynolds numbers higher than $\mathrm{Re}=37$ faces two major
problems: (a) the triggering of numerical instabilities due to large
inter-species coupling and smallness of relaxation time values;
and (b) the approach to the compressible limit, whose macrodynamic
behaviour for pure phases cannot, by construction, be correctly
described by our method. On the other hand, there is still
scope to achieve Reynolds numbers smaller than $\mathrm{Re}=0.18$ in
search of the end of the crossover to $t^{1/3}$. Closeness to
the miscibility threshold may make this endeavour difficult, as it is 
reached for characteristic times ca. $T_0=1.43\times10^{8}$.

Our results clearly challenge the claim that a domain growth
linear with time is universal for all models of phase separating
fluids sharing similar values of $L_0$ and $T_0$ since we obtained
excellent collapse of scaled structure functions yet our domain
growth exponents are in the crossover region between diffusive and
hydrodynamic viscous regimes.

The properties of this binary immiscible fluid model are important for
underpinning the more complex domain growth observable in ternary
amphiphilic (oil/water/surfactant) fluids we expect to report in
forthcoming publications.

\section{Acknowledgments}
This work was supported by EPSRC grants GR/M56234 and RealityGrid
GR/R67699 which provided access to Cray T3E-1200E, SGI Origin2000 and
SGI Origin3800 supercomputers at Computer Services for Academic
Research (CSAR), Manchester University, UK, and by the Center for
Computational Science, Boston University, USA, through a collaborative
project to access their several SGI Origin2000 platforms. We also
thank the Higher Education Funding Council for England (HEFCE) for our
on-site 16-node SGI Onyx2 graphical supercomputer. We wish to thank Dr
Hudong  Chen and Dr Peter J Love for useful discussions, Mr Jonathan
Chin for implementing the parallel code, and Dr Keir Novik for
technical assistance. NGS also wishes to thank Dr Ignacio
Pagonabarraga for useful discussions, and Prof David Jou, Prof Jos\'e
Casas-V\'azquez and Dr Juan Camacho at the Universitat Aut\`onoma de
Barcelona, Spain, for their support.

\end{document}